\documentclass[aps,prb,longbibliography,twocolumn,superscriptaddress]{revtex4-2}
\usepackage{amsmath,amssymb,bm,graphicx,color,bbold,hyperref,keyval,url,latexsym,dsfont}
\usepackage{xcolor,mathtools}
\usepackage[normalem]{ulem}
\usepackage{CJK}
\usepackage{float}
\usepackage{comment}
\usepackage{tikz}
\usetikzlibrary{shapes.misc}
\usepackage{physics}

\begin{document}

% ============================================================================================
\title{Quantum Hall to Chiral Spin Liquid transition in a Triangular Lattice Hofstadter--Hubbard Model}
% ============================================================================================

% ============================================================================================
\author{Cesar A. Gallegos}
\affiliation{Department of Physics and Astronomy, University of California, Irvine, California 92697, USA}
\author{Rafael M. Magaldi}
\affiliation{Department of Physics and Astronomy, University of California, Irvine, California 92697, USA}
\author{Andrew Millis}
\affiliation{Center for Computational Quantum Physics, Flatiron Institute, New York, New York 10010, USA}
\affiliation{Department of Physics, Columbia University, New York, NY 10027, USA}
\author{Steven R. White}
\affiliation{Department of Physics and Astronomy, University of California, Irvine, California 92697, USA}
% ============================================================================================
\date{\today}%{October 24, 2025}%

% ============================================================================================
\begin{abstract}
We investigate  the weak interaction integer quantum Hall (IQH) phase, the intermediate interaction phase identified as a chiral spin liquid (CSL) and the transition between them in the triangular lattice Hofstadter--Hubbard model at a density of one electron per site in an orbital magnetic field corresponding to one-quarter flux per plaquette. 
Our primary tool is the finite system density matrix renormalization group (DMRG) method with both interaction-strength scan and fixed interaction techniques for cylinders of circumference 3, 5, and 7 and lengths up to 240. 
For the IQH phase, we use single particle exact diagonalization to clarify finite size effects, including an excess charge on the edges of our cylinders, and the limitations of  entanglement spectra degeneracies on small circumference cylinders. 
For both phases, we use DMRG to study the entanglement spectra, the entanglement entropy, and the effect of flux insertion on charge and spin pumping, all of which show key differences between the two phases.  
To study the transition, we use interaction-strength scans extending between the two phases, and apply a scaling data collapse of a bond-dimerization order parameter to extract critical exponents. 
We also extract critical behavior from the divergence of correlation lengths on the IQH side, measuring decay away from edges of both the dimerization order parameter and transverse edge currents. The critical behavior and exponents are consistent with an Ising transition in 1+1 dimensions.  Finally, we obtain excited states in various quantum number sectors finding that  the gap to a charge neutral momentum $\pi$ excitation corresponding to fluctuations of the dimerization order parameter closes in the vicinity of the critical point but gaps to other excitations remain large.
\end{abstract}
% ============================================================================================

\maketitle

%---------------------------------------------------------------------------
\section{Introduction} \label{sec:Section1}
\vskip -0.2cm
%---------------------------------------------------------------------------
Moir\'e materials, created for example by stacking two layers of van der Waals-bonded compounds, are of great current interest~\cite{Andrei2020NatMater,Mak2022NatNanotechnol}. The defining  feature of moir\'e systems---their vastly enlarged unit cell---not only yields flat-band, correlation-driven, and topological phenomena~\cite{Balents2020NatPhys,Kennes2021NatPhys}, but also means that magnetic fields accessible in laboratories correspond to order-one flux per unit cell, making the Hofstadter physics of both interacting and noninteracting systems experimentally accessible~\cite{Zang21}.  As emphasized recently by Kuhlenkamp et al.~\cite{Knap2024}, this  suggests that the triangular-lattice Hofstadter–Hubbard model is a natural minimal setting in which orbital field-driven band topology and Coulomb repulsion compete on comparable scales. 
Of particular interest is the possible emergence of a chiral spin liquid (CSL), a topologically ordered state with fractionalized excitations and robust charge-neutral edge modes that may exist at carrier concentrations of one electron per moir\'e unit cell as an intermediate phase between a small $U$  weakly interacting integer quantum Hall (IQH) phase and a conventional topologically trivial large $U$ antiferromagnet. Originally proposed by Kalmeyer and Laughlin~\cite{Kalmeyer87}, the CSL has since been identified as a ground state phase in triangular-lattice spin models with ring exchange~\cite{Cookmeyer2021CSL}. An important topic of recent research is the possible existence of CSL phases in fermionic models~\cite{szasz2020qsl,Knap2024,divic2024csl}.    It is important to note that the CSL state requires broken time reversal symmetry--either spontaneously  or (as in this paper) explicitly. 

The triangular Hofstadter-Hubbard model is described by an electron kinetic energy parametrized by a hopping $t$ and an on-site interaction $U$. In this paper we consider only half-filling,   one electron per site. In zero applied magnetic field, at small $U$ the ground state is believed to be a Fermi liquid metal; at very large $U$ the ground state is a $120^\circ$ antiferromagnetic insulator. However, for intermediate $U$ values, the two dimensional model cannot be solved on large enough systems to provide unambiguous evidence of spin liquid behavior. Many authors therefore have used density matrix renormalization group (DMRG) methods to study quasi one-dimensional versions of the triangular lattice Hubbard model.    At zero magnetic field, the nature of an intermediate phase with $U \sim 5-8$ has attracted much attention, but whether it is a CSL or instead a time-reversal symmetric phase has been controversial~\cite{szasz2020qsl,Zhu24,Peng21,Tocchio21}.

A magnetic field that couples to the electrons' orbital motion, but not their spin,  leads to richer physics. At small $U$
the ground state is an IQH topological phase characterized by a nonzero Chern number.  At large $U$ a Schreiffer-Wolff transformation generates at order $t^2/U$ a nearest neighbor isotropic Heisenberg exchange favoring the topologically trivial $120^\circ$ magnetic state, and at order $t^3/U^2$  a time reversal breaking three spin ring exchange interaction that can energetically favor a CSL at intermediate $U$~\cite{Zang21,Knap2024}. Kulenkamp et al. provided evidence for this CSL in the electronic model using infinite DMRG (iDMRG) on width 6 cylinders, combining edge-mode counting with flux-insertion to show vanishing charge but quantized spin pumping at $U$ larger than a critical value $U_c \sim 11$. This phase was found to  persist to quite high $U$,  $U \sim 20$ or higher before a second transition to the $120^\circ$ magnetic state.  On the circumference  $4$ and $6$ cylinders studied by Kuhlenkamp et al., the change from the IQH to apparent CSL phase was associated with a weak maximum in a correlation length but no definite evidence for a bulk transition was found.  

Divić et al.~\cite{divic2024csl} uncovered a nonsymmorphic glide particle-hole symmetry relevant to odd-circumference cylinders. Spontaneous breaking of this symmetry leads to variations in bond strength between even and odd columns of bonds; the  bond strength variation defines a  charge conserving order parameter. For cylinders of circumference $L_y=3$ and $5$, also using iDMRG, Divić et al. found an apparently continuous transition marked by the onset of this order parameter.  The transition was associated with a diverging correlation length  and with a change in topology of edge states as determined from entanglement spectra. Similar bond strength alternation on odd cylinders has been observed in the spin liquid phase of the triangular lattice $J_1$-$J_2$ Heisenberg model and its $XXZ$ extension~\cite{Zhu2015J1J2TL, XXZTL2025}.  Divic et al. suggested that the critical point they identified on finite radius cylinders would develop into a 2+1 dimensional quantum topological critical point  as the cylinder circumference diverged. 

%---------------------------------------------------------------------------
\begin{figure}[t]
\includegraphics[width=\linewidth]{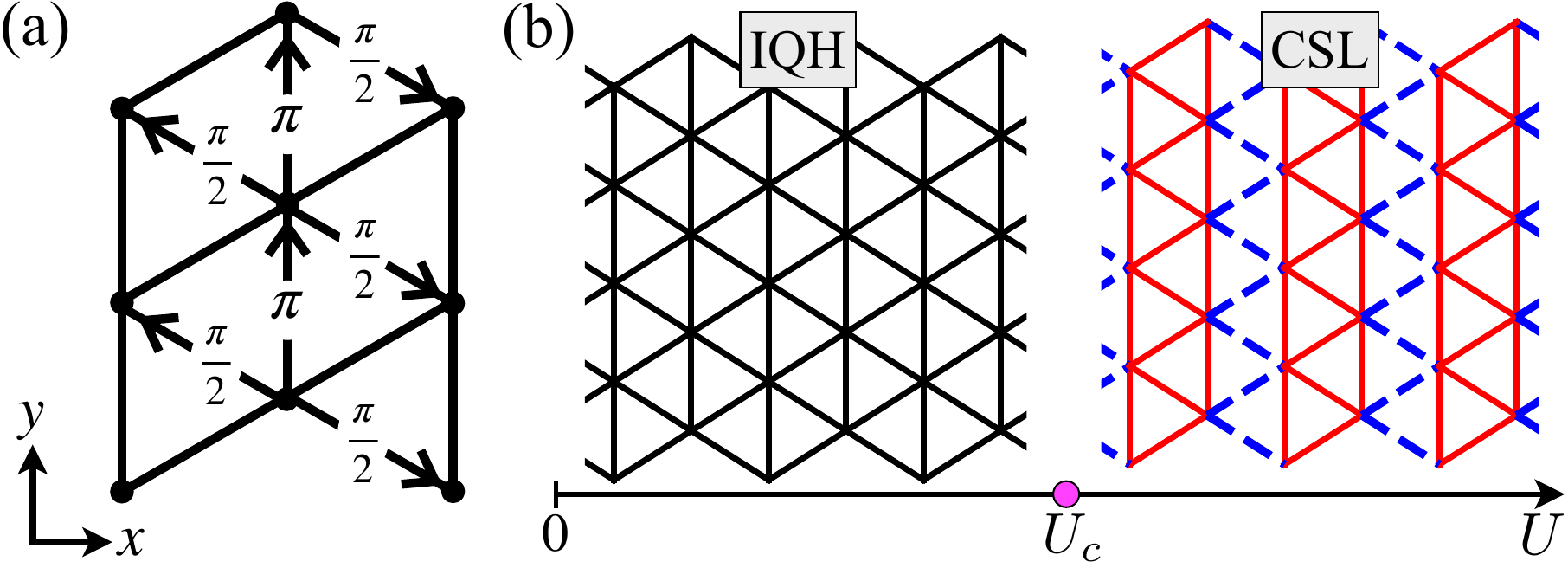}
\vskip -0.3cm
\caption{(a) Triangular lattice on a YC$3$ cylinder, with periodic boundary conditions in the $y$‑direction. Peierls hopping phases, where nonzero, are shown.  A standard cylinder starts on a ``lower" $y$-column without the extra $\pi$ phases. (b) Schematic phase diagram of the model \eqref{eq:HHmodel} at half-filling, showing the transition point $U_c$ between the integer quantum Hall (IQH) and chiral spin liquid (CSL) phases. The width of the lines are proportional to the nearest-neighbor spin correlations $\langle {\bf S}_i \cdot {\bf S}_j\rangle $ on the YC$5$ cylinders. In the CSL, a uniform offset of $–0.149$ has been subtracted from the correlations to highlight the order parameter; solid red (dashed blue) bonds indicate correlations below (above) this offset.}
\label{fig:LatticePhases}
\vskip -0.3cm
\end{figure}
%---------------------------------------------------------------------------

In this paper we revisit this transition using finite DMRG, which offers several techniques not available in iDMRG. Most importantly, scan techniques~\cite{chepiga2021scan} in which  $U$ is varied along the cylinder length allow us to observe the continuous onset of order and extract critical exponents $\beta$, and $\nu$ through data collapse. We also obtain singlet, triplet, and single and double particle gaps in the critical regime, finding that only the singlet gap closes.  We obtain entanglement entropies and spectra, and study the effect of flux insertion on charge and spin pumping, which further support the proposed existence of a CSL phase. 

We study the Hubbard model on a triangular lattice in which electrons can hop between nearest neighbor sites with amplitude $t$ and are subject to  a uniform magnetic flux $\Phi = \pi/2$ per triangle imposed by modulating the hopping amplitude between sites $i$ and $j$ with a Peierls phase $A_{ij}$:
\begin{equation}
H = -\sum_{\langle ij\rangle,\sigma} t\,e^{iA_{ij}} c^\dagger_{i\sigma} c_{j\sigma} + U \sum_i n_{i\uparrow}n_{i\downarrow}.
\label{eq:HHmodel}
\end{equation}
We have set the Zeeman coupling to zero, and set $t=1$ as the energy scale throughout. The choice of Peierls phases is shown in Fig.~\ref{fig:LatticePhases}, and is the same as in Divic et al.~\cite{divic2024csl}. This gauge choice makes the system translationally invariant in the $y$ direction.
At $\Phi = \pi/2$, the noninteracting band structure hosts Chern bands with a gap around zero energy, so that at half-filling and $U$ less than a critical value $\approx 11t$, the ground state is an integer quantized Hall (IQH) state with hall number $\nu=1$ for each spin. At very large $U$ the ground state is expected to be a topologically trivial antiferromagnet. It is believed~\cite{Knap2024,divic2024csl} that a CSL phase is the ground state for a range of intermediate $U$.

We use finite system DMRG to study cylinders  of finite length obtained by imposing periodic boundary conditions in the $y$ direction and open boundary conditions in the $x$ direction; see Fig.~\ref{fig:LatticePhases}. We  consider cylinders with circumferences $L_y=3,\ 5$, and $7$ (depending on the quantities studied), and lengths $L_x$ up to 240. We employ both ``scans" (in which a Hamiltonian parameter varied along the length of the cylinder) and ``nonscans” (all parameters fixed) with  bond dimensions typically up to $\chi=6000$ on YC5 to ensure good convergence with truncation errors of $\mathcal{O}(10^{-6})$, and up to $\chi=16000$ on YC7. 
We utilize several different specific techniques, such as scan scaling collapse to study critical exponents at the transition, and restricted sweeping for excited states, which are detailed in subsequent sections.
We first consider the IQH phase in Section~\ref{sec:Section2}, then the CSL phase in Section~\ref{sec:Section3}, the transition in Section~\ref{sec:Section4}, and excited states in Section~\ref{sec:Section5}.  In Section~\ref{sec:Section6}, we discuss our results and conclusions.

%---------------------------------------------------------------------------
\section{IQH phase} \label{sec:Section2}
%---------------------------------------------------------------------------
At $U=0$, a single particle diagonalization can be used to understand the properties of the IQH state on a finite cylinder for a wide range of lengths and circumferences.  This is very useful for understanding the finite size effects, including edge modes, on the finite small circumference  cylinders accessible to DMRG.  

\begin{figure}[t]
\includegraphics[width=\linewidth]{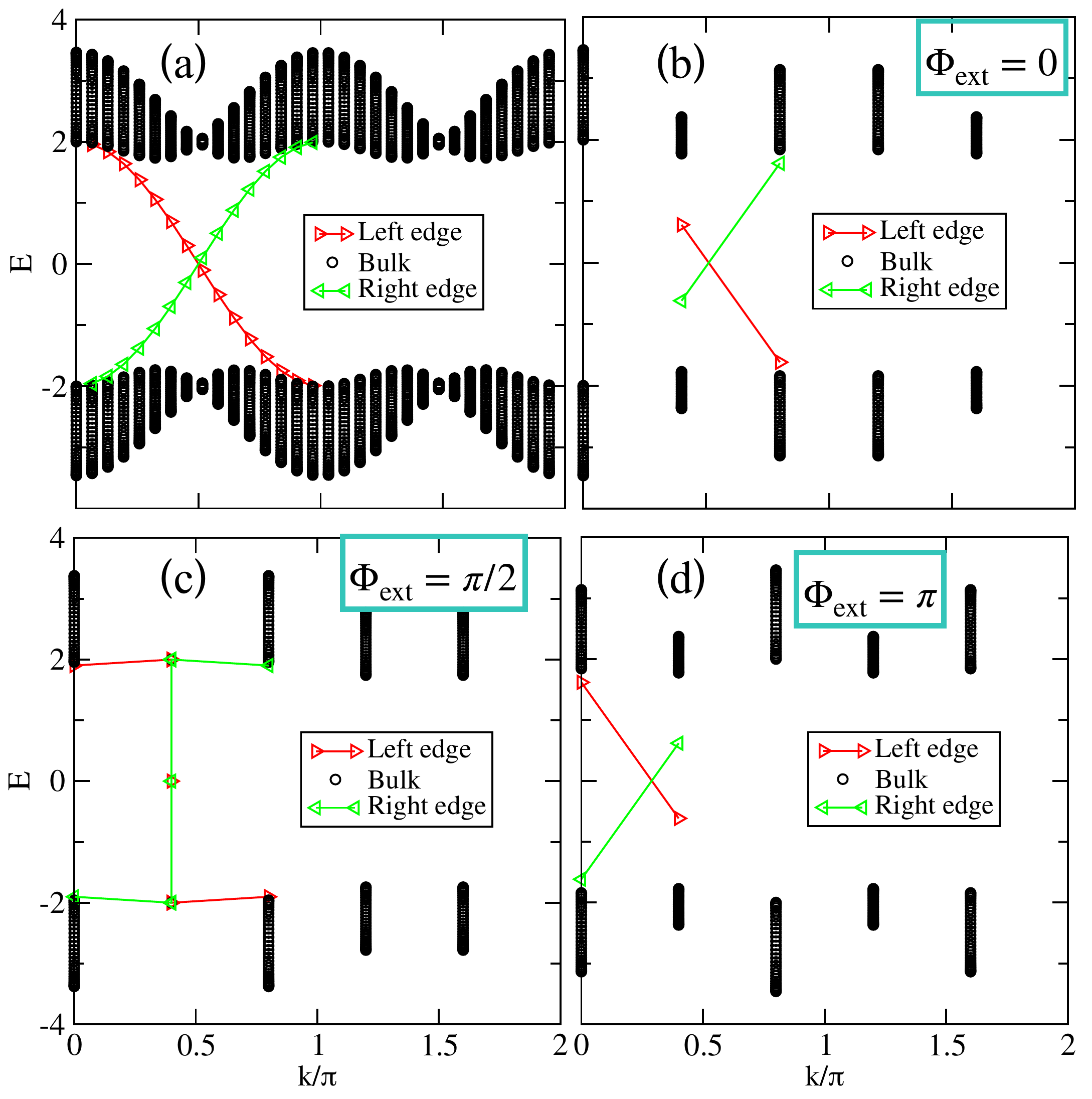}
\caption{Single particle energy levels for the model at $U=0$. (a) $64\times31$ open cylinder, showing bulk states (black circles) and left (red triangles) and right (green triangles) chiral edge bands.  (b-d) $64\times5$ cylinders, with three values of an external flux along the axis of the cylinder. For $\Phi_{\rm ext}=\pi/2$, every energy level is doubly degenerate.}
\label{fig:FigU0}
\end{figure}

In Fig.~\ref{fig:FigU0}(a) we show results obtained for a large system of spinless fermions that reveal the expected integer quantum hall physics. Translation in the $y$ direction is a symmetry of the system, so all states can be labeled by their $y$ momentum. The states obtained were also  labeled as left or right if more than 90\% of their probability was in the left or right half of the system; otherwise, they were labeled ``bulk". In addition to the bulk upper and lower bands, separated by a large gap of about $4$, we see a discrete set of states corresponding to  two chiral edge modes. In addition to the chirality, the edge modes are identified by the spatial localization of their charge density (not shown).  In such a large system, these edge modes are also readily identified by degeneracies in the entanglement spectrum (see below).  In panels (b)-(d), we show results obtained on a  $64\times5$ system that is readily accessible to DMRG, threaded by different fluxes applied along the cylinder axis.  Panel (b) shows the energies for the case of  zero threading flux. We see that only two left and two right edge modes are identifiable and distinct from the bulk. The small number of easily identifiable edge states limits the ability of entanglement spectra to unambiguously identify the edge modes.  Note also that  the left and right edge modes are not degenerate in energy; the left-right asymmetry also applies to the bulk modes although this is not readily apparent in Fig.~\ref{fig:FigU0}. The lack of both degeneracy and left-right reflection symmetry is a finite size effect arising from the combination of the glide particle-hole symmetry and  the open boundary conditions.

%---------------------------------------------------------------------------
%---------------------------------------------------------------------------

We are interested in the half-filled system, obtained by filling all of the states up to $E=0$, the chemical potential for half-filling. This ground state is non-degenerate due to the finite size.  On a small width system, there is a noticeable difference (see section~\ref{sec:Section4}) in the occupancies of odd and even columns, but this
disappears rapidly with width.  However, the system has an excess charge density on the right edge (and a deficit on the left) which does not go away with increasing width.  This excess charge persists throughout the IQH phase and is important in interpreting our DMRG results.

The magnitude of the excess charge can be varied by threading the cylinder with a flux $\Phi_{\rm ext}$.   The flux is conveniently represented by additional Peierls phases from a uniform vector potential in the $y$ direction, which does not break $y$ translational symmetry.  In this representation the allowed momenta do not change. The vertical bonds pick up an additional phase of $\Phi_{ext}/L_y$, while the extra phase of the diagonal bonds is smaller by a factor of two (i.e. $\sin \pi/6$).  If the threading flux is $\Phi_{\rm ext}=\pi/2$, the reflection symmetry is restored,  and, as shown in (c), every energy level is doubly degenerate. Also, if one leaves the two $E=0$ states empty (or filled), the total density is reflection symmetric. At a flux of $\Phi_{\rm ext}=\pi$, as shown in (d),
the density is exactly left-right reflected relative to $\Phi_{\rm ext}=0$.  Note that which side the excess charge density is on alternates between cylinder  circumferences $4n+1$ and $4n+3$.

In a $\nu=1$ quantized Hall state, changing the flux by $2\pi$ pumps one electron; thus adiabatically increasing the flux from $0$ to $\pi$ pumps $1/2$ electron from the right edge to left.  The interchange symmetry that we see between the edge states at $0$ and $\pi$ then says that we have an excess charge density of $\pm 1/4$ at an edge.
These do not correspond to extra fractionalized particles on the ends, since the  excess charge changes in magnitude as $\Phi_{\rm ext}$ is varied.  Making the system spinful, as in the Hubbard model of interest, means that the excess charge is doubled, to $\pm 1/2$ at each edge.

The sizable bulk gap makes the $U=0$ IQH phase stable  from $U=0$ to a critical value $U_c\approx 11t$.  In this phase, the Chern number is unchanged and so the charge pumping is unchanged, the  excess charge remains fixed in magnitude (for given threading flux), and the edge states remain. However, for $U$ larger than a critical value we expect a Mott insulating phase or phases, in which charge is localized,  the IQH state is destabilized and threading flux does not lead to charge pumping. 

%---------------------------------------------------------------------------
\section{Chiral Spin Liquid Phase} \label{sec:Section3}
%---------------------------------------------------------------------------

%---------------------------------------------------------------------------
\subsection{Overview and DMRG initialization} \label{sec:Section3A}
%---------------------------------------------------------------------------
For intermediate values of $U$, on odd cylinders, a distinct {non-IQH phase  is present. This phase is a Mott insulator, in which charge fluctuations are suppressed, and is dimerized  so that the glide particle-hole symmetry is broken. It is also believed to be a finite-cylinder version of the CSL phase, so we will refer to it as the CSL. Our analysis of the CSL in this section is organized by the key computational analysis techniques used to understand this phase and its relation to a CSL as one approaches 2D. 

We tried multiple initializations for the DMRG calculations in order to ensure that we did not obtain metastable states rather than the ground state.  Random initial states were a very poor starting state, leading to localized defects which were difficult to remove. However, starting with the exact $U=0$ single IQH state using the efficient algorithm of Fishman and White~\cite{Fishman2015GaussianMPS}, and then performing sweeps as $U$ was successively raised always found the ground state. In addition, starting with a N\'eel state, or starting with the $U=0$ state but immediately using the final $U$ in sweeps, were both normally successful.  Some care was needed concerning localized effective  spin‑$1/2$ “spinons” at the edges; these tended to be metastable but not present in the ground state.  The lowest final energy could always be used to identify a metastable state. For the case of scan calculations as $U$ is varied through the transition, some extra complications appear and are discussed in the  Section~\ref{sec:Section4}.

%---------------------------------------------------------------------------
\subsection{Entanglement Spectrum} \label{sec:Section3B}
%---------------------------------------------------------------------------

%---------------------------------------------------------------------------
\begin{figure}[t]
\includegraphics[width=0.85\linewidth]{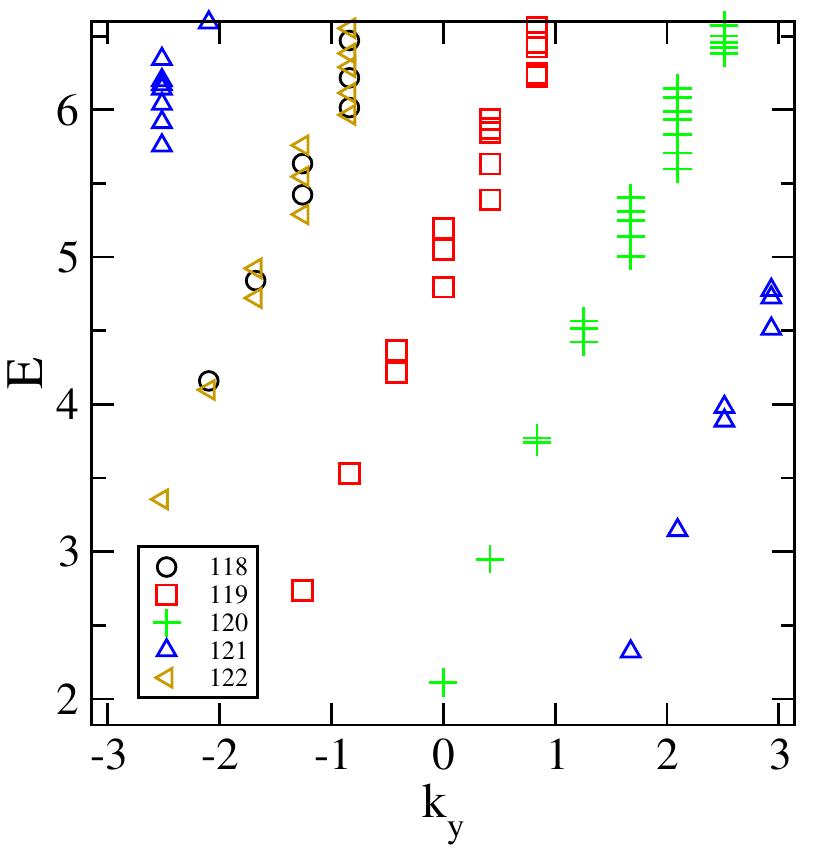}
\caption{Entanglement spectrum for different particle number sectors at $U=0$ for a $32\times15$ spinless fermion system divided between columns 16 and 17.  The legend shows the number of particles on the left half of the system.}
\label{fig:FigU0es}
\end{figure}
%---------------------------------------------------------------------------

The entanglement spectrum (ES) can be used to identify topological phases by characterizing their edge states. An ES is associated with a bipartition of the lattice, and is defined as the set of logarithms of eigenvalues  of the reduced density matrix $\left\{E_i\right\}=-\left\{\log(p_i)\right\}$, where $p_i$ is a Schmidt-decomposition probability, specifically an eigenvalue of the left-side reduced density matrix. Following the work of Li and Haldane~\cite{LiHaldane2008}, a substantial literature has shown that the low-lying eigenvalues are in one-to-one correspondence with low-lying edge modes.  Here we consider only cuts between columns, in which case the $y$ momentum and total $z$ component of spin, and total charge are good quantum numbers, which we use to label the states. The detailed construction of the momentum labeling procedure is provided in the Supplemental Material. 
 
We first consider a spinless version of our system at $U=0$, as in Fig.~\ref{fig:FigU0}. In Fig.~\ref{fig:FigU0es} we show results for a $32\times15$ system.  In this case we can clearly see a degeneracy pattern for $N=120$, which is the most probable particle number, obeying one particle per site. Specifically, we see the sequence 1,1,2,3,5,7,..., which corresponds perfectly  to a single linearly dispersing chiral mode in a finite-size system. The pattern is also visible for $N \pm 1$. To understand the pattern, note that the ground state, corresponding to filling all of the edge states up to the chemical potential,  is nondegenerate, as is the first excited state (corresponding to moving one particle from the highest filled edge state to the lowest unfilled chiral edge state). The next excitation is two-fold degenerate, reflecting the two possible excitations (degenerate for a linear spectrum) of moving a particle from the highest filled to the second lowest empty state, or moving one particle from the second highest filled to the lowest empty state.   The pattern for a spinful system would be 1,2,5,10,20,... It is important to note that degeneracies are not perfect because the finite size of the system means that the energy gaps between successive states are not infinitesimal, so nonlinearities in the energy spectrum may occur.

%---------------------------------------------------------------------------
\begin{figure*}[t]
\centering
\includegraphics[width=\textwidth]{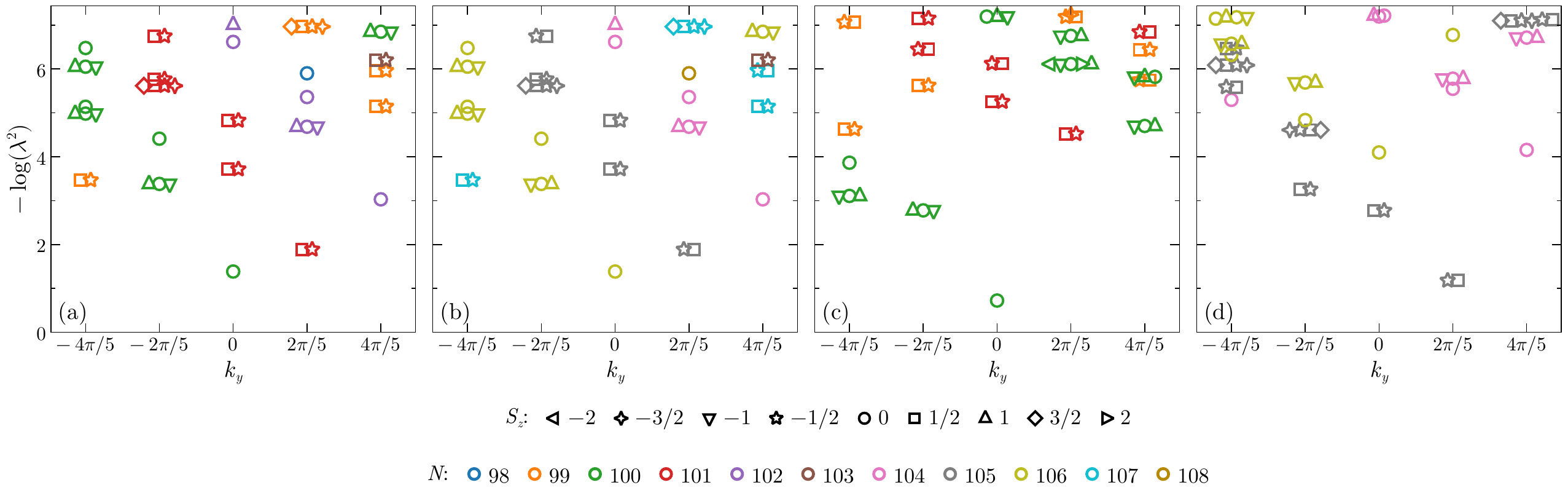}
\caption{Entanglement spectrum versus transverse momentum \(k_y\) for a YC5 cylinder of length \(L_x=80\) at bond dimension \(\chi=2000\). Panels (a,b) show \(U=6\); panels (c,d) show \(U=15\). For each \(U\) we compare adjacent bipartitions taken immediately to the right of columns 20 and 21 (even/odd cuts).  At \(U=6\) the cuts immediately to the right of columns 20 (panel a) and 21 (panel b)  are essentially identical up to a relabeling of particle number.  At \(U=15\) we see that the cut immediately to the right of column 20 (panel c) has a non-degenerate entanglement spectrum ground state and an approximately three-fold degenerate next lowest state, differing  markedly from the cut immediately to the right of column 21 (panel d) which has a two fold degenerate entanglement ground state and a two-fold degenerate next lowest state. Marker color encodes the particle number \(N\) on the left of the cut and marker shape encodes \(S_z\) of the left Schmidt sector (see legends). Only the lowest 50 entanglement levels are shown.}
\label{fig:ESwithKy}
\end{figure*}
%---------------------------------------------------------------------------
 
With DMRG, for the ES, we are limited to width 5, where only a small number of levels are available to track degeneracies and the nonlinearies that break the theoretically expected degeneracies are therefore large. Nevertheless, the ES still provides valuable information distinguishing the IQH and CSL phases. 

Even and odd cuts are distinguished by the number of lattice sites to the left of the bipartition.  All spectra on YC-5 cylinders quoted below are taken at $\ell=L_x/4$ and $\ell=L_x/4 + 1$, that is, around one‑quarter of the cylinder length from the left edge.  Positioning the cut away from the midpoint reduces the cost of computing the momentum labels, because the translation operator $T_y$ then acts on the shorter half‑cylinder.  A systematic scan of $\ell$ (see Supplemental) shows that changes in the low‑lying entanglement energies are minimal once the cut is sufficiently away from the left edge, so the off‑center choice introduces no measurable bias while greatly alleviating the numerical workload.

For large $U$, a Mott-Hubbard picture would say that the most probable Schmidt states would have the number of electrons equal to the number of sites. This implies that the dominant spin multiplets should be either integer or half-integer for the even and odd cuts. In addition, given both the spin liquid nature of the CSL and its column dimerization, it is natural to consider a near-neighbor dimer gas picture. In this case, we expect that if the entanglement cut does not break a dimer  (i.e. if the number of sites to the left [and right] of the cut is even), then we expect a unique $S=0$ entanglement ground state and a three-fold degenerate next entanglement eigenstate. However, if the entanglement cut breaks a dimer then we expect a spin doublet to be the most probable ground state and most probable first excited state. Exactly this behavior is seen in Fig.~\ref{fig:ESwithKy}, panels (c) and (d).  However, the excess edge charge of $\frac{1}{2}$ in the IQH phase means that the most probable post-cut states are spread equally between two adjacent particle numbers.   Remarkably, this makes the even and odd column ES essentially identical (except for particle number relabelings) as seen in 
Fig.~\ref{fig:ESwithKy}, panels (a) and (b).  This is the key signature of the IQH phase. On this width-5 system one can see the start of the expected degeneracy pattern ($1,2,5,10...$) but only in the values (1,2) within the first 50 states, and the degeneracy is imperfect. Note that we are ignoring the spin-degeneracies; for the degeneracy 2 at $k_y=-2\pi/5$, there are four low-lying symbols in each panel, but three form an exact spin triplet so we count this as one state. In the CSL case, each edge has one chiral mode, and one expects degeneracies in the pattern ($1,1,2,3,5,\ldots$).  For the odd cut (d) in Fig.~\ref{fig:ESwithKy}, one can see the pattern for $N=105$ out to 4 values, ($1,1,2,3$).

To summarize: the entanglement spectra clearly show that both the smaller-$U$ and larger-$U$ phases have edge states within the bulk gap, but the associated pattern of degeneracies differs from the smaller-$U$ to the larger $U$ phases. The differences in degeneracies and presence or absence of variation between even and odd cuts clearly distinguish the phases and support their identification as IQH and CSL. The results are in agreement with the ES obtained in Ref.~\cite{divic2024csl}, but the detailed quantum number assignments and comparison to the $U=0$ case provide additional insights.

%---------------------------------------------------------------------------
\subsection{Entanglement Entropy} \label{sec:Section3C}
%---------------------------------------------------------------------------
Complementary information comes from the von Neumann entanglement entropy (EE).   For a bipartition \(A|B\) across the central cut, we write the Schmidt decomposition
\begin{equation}
|\psi\rangle=\sum_{\alpha}\lambda_{\alpha}\,|\alpha_A\rangle|\alpha_B\rangle,
\qquad
\sum_{\alpha}\lambda_{\alpha}^{2}=1 .
\end{equation}
The von Neumann entropy is then
\begin{equation}
S(A)=-\sum_{\alpha}\lambda_{\alpha}^{2}\ln \lambda_{\alpha}^{2},
\end{equation}
i.e., \(S=-\mathrm{Tr}\,\rho_A\ln\rho_A\) with eigenvalues of \(\rho_A\) given by \(\{\lambda_\alpha^{2}\}\). We consider two aspects of $S$:  (1) What is its behavior as a function of system length and versus the parity of the partition in the two phases and at the critical point? For this we focus on YC3. (2) Can we extract a topological EE term from our data, in the two bulk phases?

%---------------------------------------------------------------------------
\begin{figure}[t]
\centering
\includegraphics[width=0.9\linewidth]{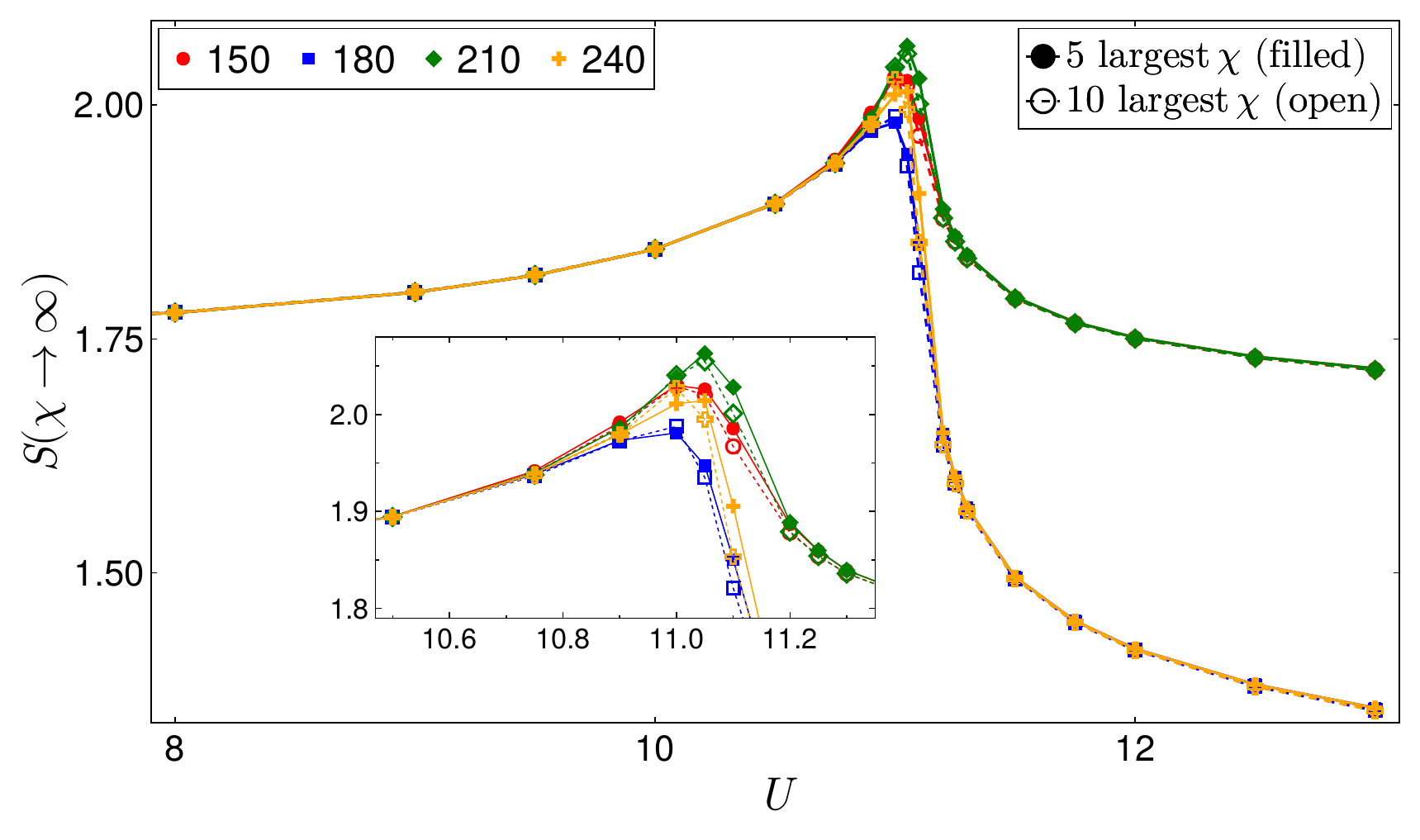}
\caption{Extrapolated entanglement entropy $S$ on YC3 cylinders versus $U$. We compare two extrapolation schemes, retaining the largest 5 (solid lines, filled markers) and the largest 10 (dashed lines, open markers) bond dimensions. Cylinder lengths $L_x$ are shown in the upper left legend, and the cut is always in the middle. At larger $U$, the curves split into two branches depending on the cut parity. Inset: zoom of the transition region.}
\label{fig:SvN}
\end{figure}
%---------------------------------------------------------------------------

In Fig.~\ref{fig:SvN}, we measure the entropy at the center of the YC3 cylinder for lengths $L_x=150$, $180$, $210$ and $240$, so that the center cut gives an even cut for half the cylinders ($L_x=180$ and $240$) and an odd cut for the other half ($L_x=150$ and $210$). We follow the entropy evolution as a function of bond dimension $\chi$, and extrapolate its value by linear fits in $1/\chi^p$, where the exponent $p$ is optimized over all datasets (see Supplemental for details). In the analysis we retain either the last five or the last ten $\chi$ values in the fit, providing a measure of the robustness of the extrapolation.  We see that for $U<U_c\approx 11.05t$ the entanglement entropy is independent of the location of the entanglement cut and weakly increases with increasing $U$, reaching  a maximum at $U/t\approx11.05$. For $U>U_c$, the entanglement entropy takes one of two values, depending on the system size (mod $4$).  The branch followed correlates exactly with the cut parity: systems where the left side has an odd number of sites have substantially larger $S$.  This corresponds to half-integer spin quantum numbers, and is consistent with the behavior of the ES. The larger entanglement obtained for the odd cuts is what one would expect from a near-neighbor dimer picture for the CSL, where in any dimer covering an odd cut must cut a dimer in two. 

One expects a divergence in the entanglement entropy at a continuous phase transition, with a $\ln(L_x)$ dependence on $L_x$, and the data qualitatively support this.} This result, combined with the column-parity independent ES in the IQH phase, the contrasting parity splitting in the CSL ES, the limited edge counting degeneracies, and the two‑branch structure of $S$ give solid support for a continuous transition between a IQH state and a CSL.

In two dimensional systems with topological order, the asymptotic large $L_y$ limit of the entropy law, taking $L_x \to \infty$, can exhibit a constant topological $\gamma$  term~\cite{Eisert2010EEReview}.  Here, to emphasize the large $L_y$ nature of this term, we divide the usual expression by $L_y$:
\begin{equation}
\frac{S}{L_y}=\alpha -\frac{\gamma}{L_y}+\mathcal{O}(1/L_y^2),
\label{eq:S_Ly}
\end{equation}
where  $\alpha$ is a system-dependent constant. The topological entanglement entropy (TEE) $\gamma$ is a universal constant (which may be zero), depending on the type of topological order~\cite{Kitaev2006TEE, Wen2006TEE, Jiang2012TEE}. 
In the IQH phase we expect $\gamma^{\phantomsection}_\mathrm{IQH} = 0$~\cite{Sierra2009EEIQH}, while in the Kalmeyer-Laughlin CSL, the prediction is the same as for the $\nu=1/2$ Laughlin state, $\gamma^{\phantomsection}_\mathrm{CSL}=\frac{1}{2}\log 2\approx0.347$~\cite{Kalmeyer87}. 

For $U=15$, deep in the CSL phase, the short correlation length allowed us to use modest $L_x$, and we were able to converge the DMRG adequately to extract $S$ for widths 3, 5, and 7, keeping a bond dimensions of up to $\chi=2600$, $10500$, and $16000$ states, respectively. This led to truncation errors of $\mathcal{O}(10^{-8})$, $\mathcal{O}(10^{-6})$, and $\mathcal{O}(10^{-5})$, respectively. In determining energies accurately with DMRG, we rely on the extrapolation of the energy with truncation error, where the relation is linear~\cite{SashaSteve2007PRL}. Unfortunately,  $S$ is not expect to have a linear relation with the truncation error. We tried different power-law extrapolations, which seemed to work quite well, but unfortunately they all yielded essentially the same extrapolated $S$ (within $10^{-4}$), so we do not have error estimates for the results; see the Supplemental Material for details. Similar data for the IQH phase was obtained from the single-particle diagonalization at $U=0$ (see Section~\ref{sec:Section2}). 

%---------------------------------------------------------------------------
\begin{figure}[t]
\centering
\includegraphics[width=\linewidth]{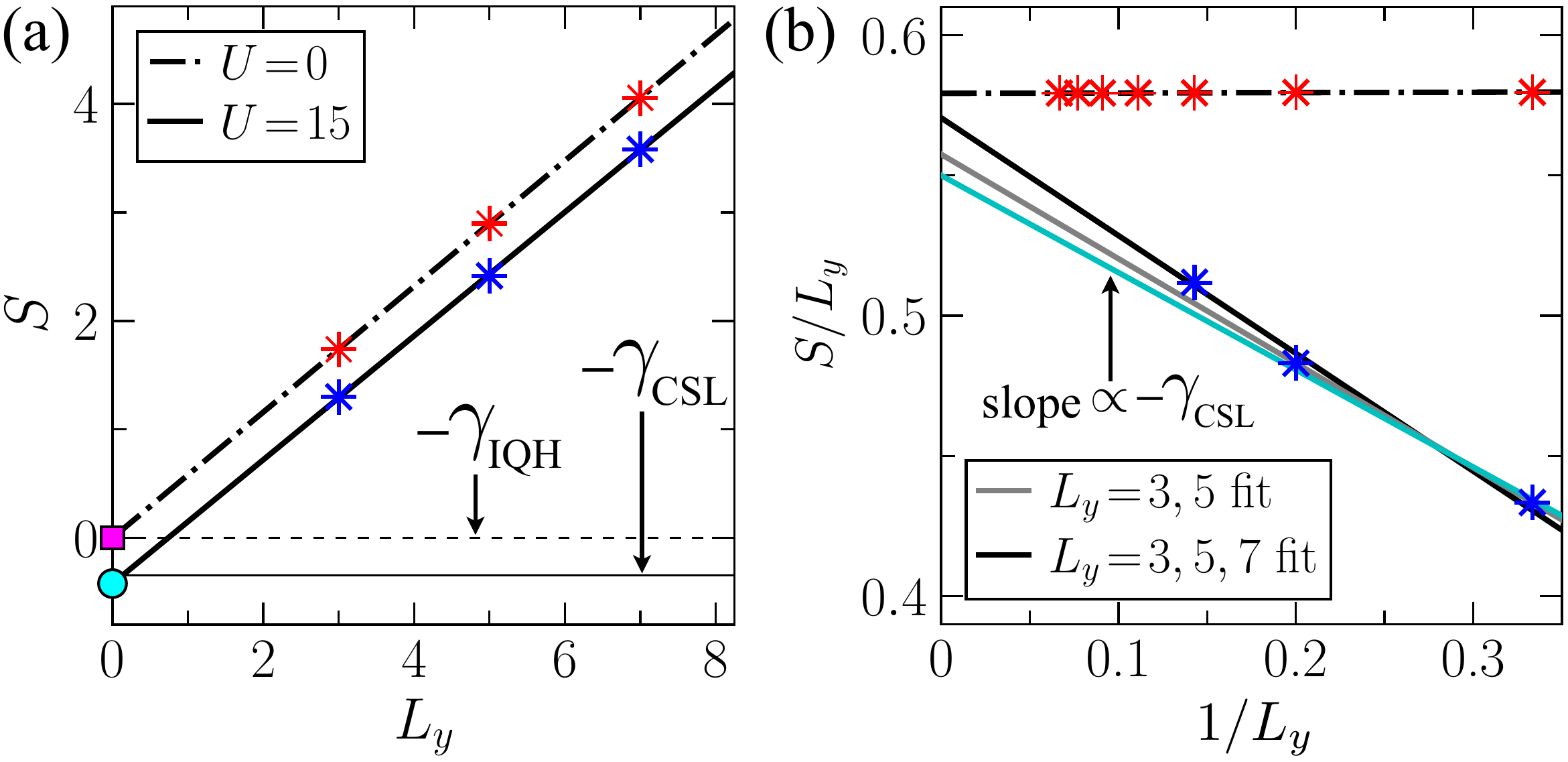}
\caption{Finite-size scaling of the entanglement entropy at $U\!=\!0$ (red stars, from single-particle diagonalization) and $U\!=\!15$ (blue stars, from DMRG). (a) $S$ vs $L_y$. Linear fits $S(L_y)=\alpha L_y-\gamma$ are shown. The extrapolated intercept at $L_y=0$ yields $-\gamma$ (filled square and circle). Horizontal lines mark the theoretical values of the topological entanglement entropy $\gamma_\mathrm{IQH}^{}=0$ and $\gamma_\mathrm{CSL}^{}=\frac{1}{2}\log 2\approx 0.347$. (b) Same data plotted as Eq.~\eqref{eq:S_Ly}. Linear fits are shown; the slope yields $-\gamma$. The light-blue line indicates the expected slope $\gamma_\mathrm{CSL}^{}$, with an arbitrary intercept.}
\label{fig:SvsL}
\end{figure}
%---------------------------------------------------------------------------

Figure~\ref{fig:SvsL} shows the EE in two complementary forms. In Fig.~\ref{fig:SvsL}(a), we show $S(L_y)$ versus $L_y$, the usual way this data is shown. The extrapolated intercept with the $y$-axis at $L_y=0$ (filled square and circle) gives $-\gamma$. The red and blue stars are the $U=0$ and $15$ data, respectively, while the corresponding linear fits are shown as dotted-solid and solid lines. The horizontal dashed and solid lines are a guide to the eye of the theoretical values of $\gamma$ for the IQH and CSL phases. However, since the expression is expected to hold in the large $L_y$ limit, it is more natural to examine the data as shown in Fig.~\ref{fig:SvsL}(b), where we plot $S/L_y$ versus $1/L_y$. Here the slope of the line corresponds to $\gamma$. In the figure, the light-blue line indicates the slope expected for the CSL phase and highlights the deviation between our numerical result and the theoretical prediction. For the IQH phase, 
the linear fit  yields an extrapolated TEE consistent with the theoretical prediction of $\gamma^{\phantomsection}_\mathrm{IQH} = 0$~\cite{Sierra2009EEIQH} within a $10^{-3}$ accuracy. 

In the CSL, we find that $\gamma\approx 0.421\ (\approx1.21\gamma_\mathrm{CSL}^{\phantomsection} )$, clearly different from the IQH result, but only in  reasonable agreement with the expected CSL result of $0.347$. The discrepancy is most likely due to incomplete convergence for the YC$7$ cylinder, where $S$  differs by $10\%$ between the last sweep and the extrapolated value. For the YC$3$ and YC$5$ cylinders, the corresponding differences are within $1\%$ and $0.1\%$, respectively. Note that if we restrict the fit to just the YC$3$ and YC$5$ cylinders, we get $\gamma\approx 0.372 \ (\approx1.07\gamma_\mathrm{CSL}^{\phantomsection} )$, in a much better agreement with the theoretical prediction.

%---------------------------------------------------------------------------
\subsection{Flux Insertion} \label{sec:Section3D}
%---------------------------------------------------------------------------
We now analyze the topological character of the candidate CSL state by studying the response of the system to fluxes $\Phi_\text{ext}$ threaded along the cylinder axis~\cite{Zaletel2014, Gong2024CSLSquare, Gong2014CSLKagome, Knap2024}.  The calculations were initialized with the converged ground state wavefunction at zero flux, and then flux was increased in steps of $\pi/8$. At each step we performed two DMRG sweeps with a bond dimension $\chi=2400$, yielding a truncation error of order $\mathcal{O}(10^{-5})$ in the $80\times5$ YC cylinders. This procedure will tend to follow a state adiabatically, but a level crossing transition may occur,  signified by a sudden drop in the energy.  In the calculations shown, this did not happen; an adiabatic path was followed. 

%---------------------------------------------------------------------------
\begin{figure}[t]
\centering
\includegraphics[width=\linewidth]{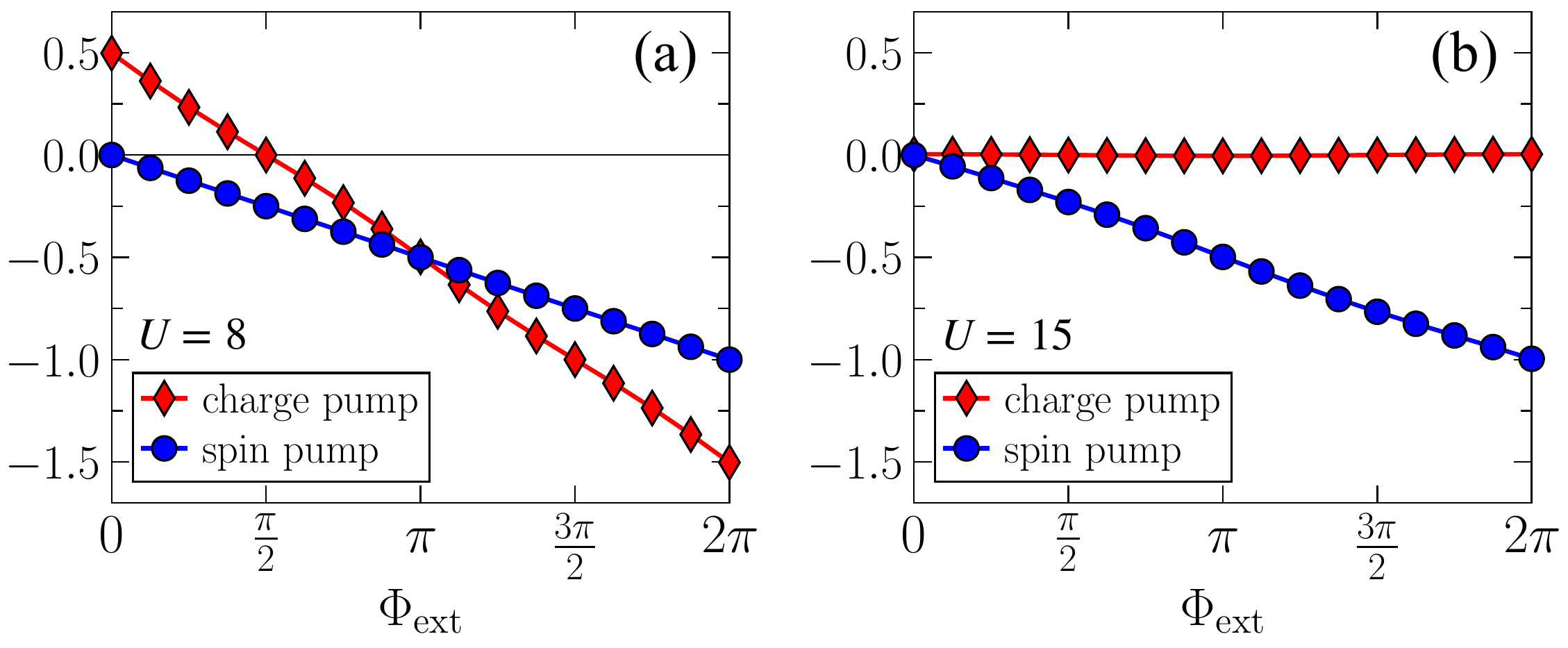}
\caption{The total charge or spin in half the cylinder under flux insertion with the same phase for spin up and down (charge pump, red symbols) or opposite phases for spin up and down (spin pump, blue symbols), respectively, for $U=8$ [(a)] and $U=15$ [(b)].}
\label{fig:pump}
\end{figure}
%---------------------------------------------------------------------------

Our Fig.~\ref{fig:pump} shows the excess charge in half of the cylinder under flux insertion with the same phase for spin up and down (charge pump, red symbols). The figure also shows the total spin in half the cylinder under flux insertion with opposite phases for the two spins (spin pump, blue symbols). For $U=8$, Fig.~\ref{fig:pump}(a), the charge pump transfers two electrons (one for each spin) as the flux increases from $0$ to $2\pi$, while the spin pump transfers a net spin of 1; both signatures are consistent with the IQH state. For $U=15$, Fig.~\ref{fig:pump}(b), the system does not pump charge, indicating a Mott-insulating state, but the spin pump still transfers one unit of spin, consistent with the $\nu=1/2$ bosonic fractional quantum Hall picture of the CSL state~\cite{Kalmeyer87,Zaletel2014, Gong2024CSLSquare, Gong2014CSLKagome, Knap2024}. In all cases, the changes in charge and spin density as the threading flux are localized with $\sim 8$ lattice sites of the ends of the cylinder. 

%---------------------------------------------------------------------------
\section{Detailed investigation of transition} \label{sec:Section4}
%---------------------------------------------------------------------------
In this section we investigate the critical behavior at the transition between the IQH and putative CSL phases.  We first present results obtained from the scan DMRG method~\cite{chepiga2021scan} in which one studies a finite system where a model parameter  is slowly varied across the system, physical quantities are computed as a function of position, and finite-size scaling is used to interpret the results, and then employ conventional DMRG methods to investigate the decay of the order parameter away from sample edges.

%---------------------------------------------------------------------------
\subsection{DMRG Scans and Scaling Collapse} \label{sec:Section4A}
%---------------------------------------------------------------------------
For our scans, the Hubbard interaction $U$ is varied linearly along the length of the cylinder with a constant gradient, 
\begin{equation}
\delta U=(U_f-U_i)/(L_x-1),
\end{equation}
where $U_i$ and $U_f$ are the values of $U$ at the ends of the cylinder. For both YC$3$ and YC$5$, we varied the scan gradient $\delta U$ in two ways: by changing the cylinder length while keeping $U_i$ and $U_f$ fixed (up to $240$ columns for YC$3$ and $80$ for YC$5$), and varying $U_i$ and $U_f$ at a fixed length.

We investigated the neutral order parameter
\begin{equation}
\mathcal{O}_x = \frac{1}{L_y}\sum_{y}(-1)^x\big( \bm{S}_{x,y}\!\cdot\!\bm{S}_{x+1,y}-\bm{S}_{x+1,y}\!\cdot\!\bm{S}_{x+2,y}\big),
\label{eq:OrderParameter}
\end{equation}
which quantifies the  dimerization of the diagonal spin–spin correlations exhibited by odd-circumference cylinders in spin-liquid states; see Refs.~\cite{Zhu2015J1J2TL, divic2024csl}. In the scans, $\mathcal{O}_x$ oscillates slightly between adjacent columns; the oscillation is due to the finite gradient $\delta U$. For typical gradients we impose, the oscillation is about an order of magnitude smaller than $\mathcal{O}_x$. For analysis, we remove this oscillation by averaging the value of $\mathcal{O}_x$ over adjacent columns, i.e., computing $(\mathcal{O}_{x}+\mathcal{O}_{x+1})/2$; see Supplemental Material for details. In scans where the range of $U$ extends well inside the IQH phase and the transition is not crossed $(U_i<U_f<U_c)$, we confirm that the order parameter decays rapidly away from the boundaries of the cylinder.

%---------------------------------------------------------------------------
\begin{figure}[t]
\includegraphics[width=\linewidth]{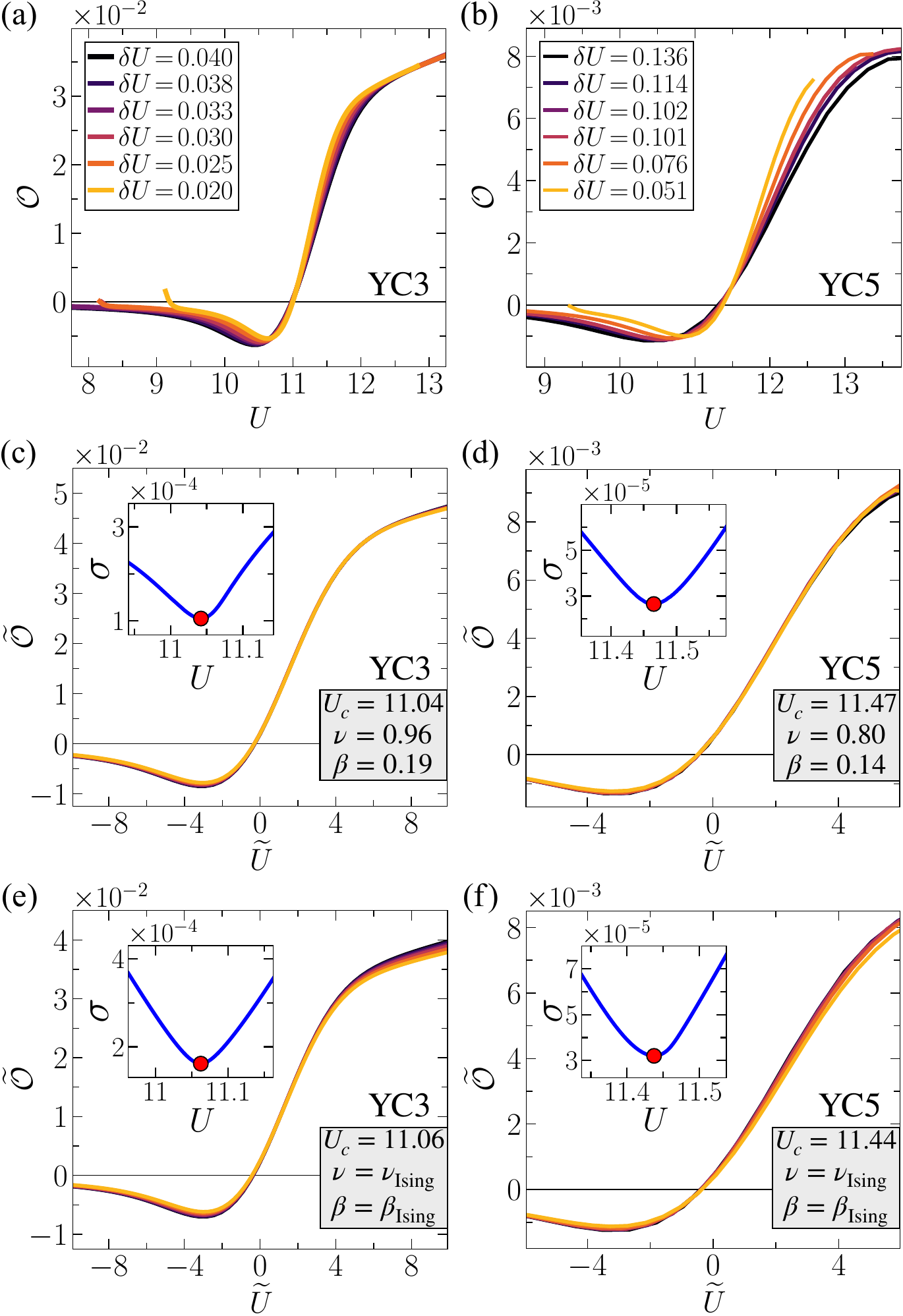}
\vskip -0.1cm
\caption{(a), (b) The DMRG scan results for the order parameter $\mathcal{O}$ on YC$3$ and YC$5$ cylinders, respectively, for various gradients $\delta U$; (c), (d) the corresponding rescaled order parameter $\widetilde{\mathcal{O}}$ with the best-fit critical point $U_c$ and scaling exponents $\nu$ and $\beta$ indicated. Insets: the standard deviation $\sigma$ of the rescaled data as a function of $U$, with the optimal $U_c$ marked by a red symbol. (e), (f) same as (c), (d), but only fitting the critical point $U_c$ and using the Ising critical exponents $\beta_\mathrm{Ising}=1/8$ and $\nu_\mathrm{Ising}=1$.}
\vskip -0.3cm
\label{fig:ScanCollapse}
\end{figure}
%---------------------------------------------------------------------------

Our Figs.~\ref{fig:ScanCollapse}(a) and (b) show the DMRG scan results for the dimerization order parameter $\mathcal{O}$ as a function of position (parametrized by value of $U$) on the YC$3$ and YC$5$ cylinders. One sees that for $U$ less than a critical value $U_c\approx11$, $\mathcal{O}$ is small and tends to zero as the gradient is decreased (the small nonzero value is a boundary effect arising from pinning of order parameter fluctuations at the sample edge), whereas for $U$ larger than the critical value, $\mathcal{O}$ tends to a  non-zero value as the gradient decreases. Interestingly, in scans in which $U_f>U_c>U_i$, $\mathcal{O}$  changes sign as a function of position, with the sign change occurring at the spatial location where $U(x)$ passes through $U_c$. This position separates a region where the order parameter is small, decaying away from the boundary, and large, a property of the infinite system. We have tried additional changes to the boundary conditions, such as removing two sites on one edge or adding/eliminating a column or an extra site at an edge (while preserving an even total number of sites). We have also tried scans such that the CSL is located between two IQH phases, by a quadratic variation of $U$ that is maximal at the center of the cluster, for example. In such cases, the system develops two sign changes: one per each IQH--CSL transition, thus providing strong evidence that in the scan method the ground state is the one with a sign change (domain wall or phase slip) in the order parameter as the interaction is scanned across the critical value.

The scans are sensitive to becoming trapped in a metastable state. For example, if one starts a scan using a non-scan DMRG ground state with a uniform large value of $U$, the system fails to develop a domain wall. This also often happens when starting from a random product state. The resulting energies are approximately $1\%$ higher than that of the true ground state. In general, we find that starting DMRG from the Néel state or from the exact solution at $U=0$ reliably yields the ground state for both the YC$3$ and YC$5$ scans.

It is also worth noting that for the YC$3$ cylinders, but not the YC5 cylinders, the domain wall/phase slip hosts a non-zero spin density. This ``spinon-like” behavior can be suppressed by modifying the boundary conditions in the YC$3$ cylinder by removing one site on both edges of the cylinder~\cite{Zhu2015J1J2TL}.

In Figs.~\ref{fig:ScanCollapse}(c) and (d), we show the same data after applying the scaling
\begin{align}
\widetilde{U}&=(U-U_c)\times \delta U^{-1/(1+\nu)},\\
\widetilde{\mathcal{O}}&=\mathcal{O}\times\delta U^{-\beta /(1+\nu)},
\end{align}
using the best‑fit values of the critical point $U_c$ and the exponents $\nu$ and $\beta$. After rescaling with these parameters, scans with different gradients $\delta U$ collapse onto a single curve regardless of the presence of the domain‑wall. For the YC3 cylinders, the estimated exponents are: $0.11 < \beta < 0.52$ (best fit $0.19$), $0.27 < \nu < 1.21$ (best fit $0.96$), and  $11 < U_c < 11.08$ (best fit $11.04$). For the YC5 cylinders, the estimated exponents are: $0.07 < \beta < 0.34$ (best fit $0.14$), $0.37 < \nu < 0.95$ (best fit $0.80$), and  $11.41 < U_c < 11.52$ (best fit $11.47$). To estimate the uncertainties on $U_c$, $\nu$, and $\beta$, we proceed by fixing two of these parameters at their best-fit values and varying the third. For each choice of the free parameter we recompute the standard deviation $\sigma$ of the collapsed data; the upper and lower bounds are then defined by the points at which $\sigma$ first reaches fifty percent above its smallest value, i.e., when $\sigma$ reaches $1.5\times$ its minimum. This procedure yields reasonable error bars that reflect how sensitively the collapse deteriorates when moving away from the optimal parameter set. We note that there is certain arbitrariness when choosing the ``window" in which $\sigma$ is computed. For our fitting procedure, we used $-5<\widetilde{U}<5$ to calculate the deviation between the curves, which is then minimized. 

For completeness, we also fixed the critical exponents to the expected ones from the Ising 1+1D universality class, $\nu_\mathrm{Ising}=1$ and $\beta_\mathrm{Ising}=1/8$, yielding a similar critical point $U_c$ and relatively similar $\sigma$ to the best-fit critical point and exponents; see Figs.~\ref{fig:ScanCollapse}(e) and (f). Interestingly, the most relevant deviations between the curves are for $|\widetilde{U}| \gtrsim 4$.

%---------------------------------------------------------------------------
\subsection{Boundary Decay} \label{sec:Section4B}
\vskip -0.2cm
%---------------------------------------------------------------------------

%---------------------------------------------------------------------------
\begin{figure}[t]
\includegraphics[width=\linewidth]{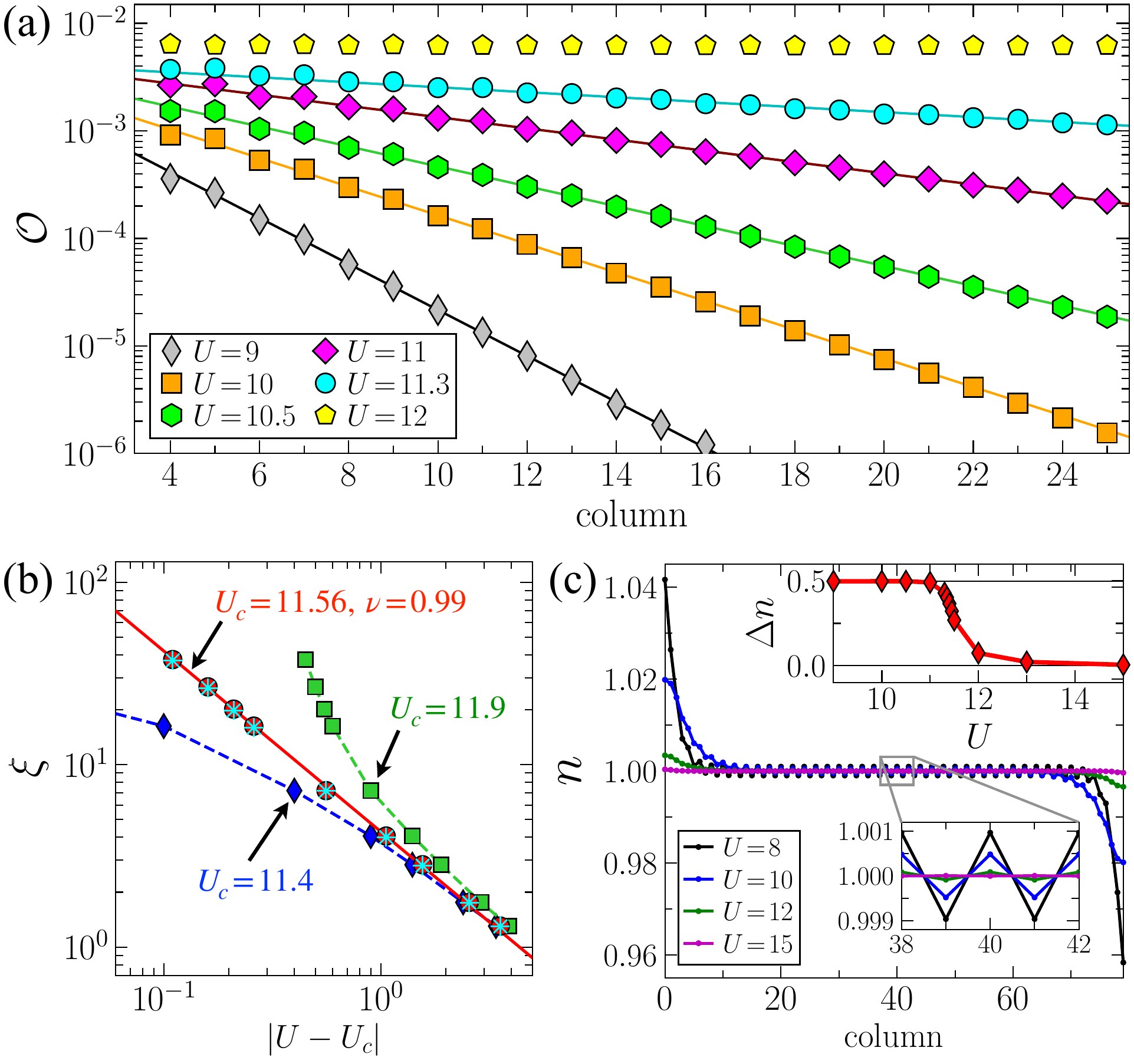}
\vskip -0.2cm
\caption{(a) Decay of the order parameter $\mathcal{O}$ as a function of distance from the edge in $80\!\times\!5$ YC cylinders, for different values of $U$. Solid straight lines in the semi‑log plot indicate the exponential fits used to extract the correlation length $\xi$ for each $U$. (b)  Log–log plot of $\xi$ versus $|U - U_c|$ for several trial $U_c$. In the curves, the choice of $U_c$ affects only the $x$-coordinate. The  red circles and line corresponds to the best fit with $U_c = 11.56$, yielding a linear dependence with slope $\nu = 0.99$. The cyan stars inside the red circles are the $\xi$ obtained from the decay of the edge currents, also using $U_c = 11.56$; see text. (c) Electron density profile $\langle n\rangle$ in $80\!\times\!5$ YC cylinders, for different values of $U$. Lower inset: zoom of the bulk charge oscillation over the indicated area. Upper inset: fractional excess number of electrons on the left half of the cylinder as a function of $U$.}
\vskip -0.3cm
\label{fig:NonScans}
\end{figure}
%---------------------------------------------------------------------------

In this subsection we use standard fixed-parameter finite DMRG calculations to investigate the decay of the order parameter from an edge pinned by the open boundary conditions. We focus on YC$5$ cylinders of length up to $80$ sites, keeping $4000$ states in all calculations. For each fixed $U$, we average $\mathcal{O}$ over each column and plot it versus the distance from the edge in Fig.~\ref{fig:NonScans}(a) on a semi‑logarithmic scale. To good approximation, the order parameter is seen to decay exponentially away from the edge $\left|\mathcal{O}(x)-\mathcal{O}(x=0)\right|\sim e^{-x/\xi}$. Straight‐line fits to the logarithm of the order parameter yield the correlation length $\xi (U)$. Since the correlation length $\xi$ diverges as $U$ approaches the critical $U_c$ according to
\begin{align}
\xi \propto |U-U_c|^{-\nu },
\end{align}
we extract $U_c$ and $\nu$ by plotting $\xi$ versus $|U-U_c|$ on a log-log scale for several trial values of $U_c$. The proper $U_c$ is identified when the data fall on a straight line; as shown in Fig.~\ref{fig:NonScans}(b), the best fit yields 
\begin{align}
U_c\!=\!11.56,\quad \nu =0.99(2),
\end{align}
consistent with the one‑plus‑one–dimensional Ising universality class expected for the scalar dimerization order parameter, where $\nu=1$. In that panel the blue and green symbols illustrate how choosing slightly smaller or larger $U_c$ visibly spoils the linear collapse, demonstrating the robustness of this approach.

We have also computed the electron currents,
\begin{equation}
J_{ij} = i \sum_\sigma \langle \left( t_{ij} e^{i A_{ij}} c_{i\sigma}^\dagger c_{j\sigma} - t_{ji} e^{i A_{ji}} c_{j\sigma}^\dagger c_{i\sigma} \right)\rangle,
\end{equation}
which in the IQH phase flow along the cylinder’s circumference and decay exponentially into the bulk, alternating in sign from ring to ring. Fitting the decay of the envelope of the oscillating current  yields the same correlation length $\xi(U)$ as obtained from the order‑parameter profiles and are shown as cyan stars in Fig.~\ref{fig:NonScans}(b), further confirming our analysis.

Finally, we study the charge density on non-scans for different values of $U$. Fig.~\ref{fig:NonScans}(c) shows the average density $n=\langle n_{\uparrow} +n_{\downarrow}\rangle $ on each column of the YC$5$ cylinders. It is clear that the excess (deficit) of charge on the left (right) of the cylinder is highly localized at the edges for small $U$ and becomes progressively flatter (eventually disappearing) as we approach the IQH-CSL transition point. It is worth noting that the small $\big[\mathcal{O}(10^{-4})\big]$ bulk charge oscillations described in Sec.~\ref{sec:Section2}, which arise from finite-size effects, vanish in the CSL phase (see lower inset). We also measure the excess edge charge $\Delta n$ on the left half of the cylinder. As shown in the upper inset, $\Delta n$ corresponds to $1/2$ in the IQH phase and it drops to zero in the CSL phase. This confirms that the higher $U$ phase is Mott-like, with no charge pumping and suppressed charge fluctuations.

%---------------------------------------------------------------------------
\section{Excited States and Gaps} \label{sec:Section5}
\vskip -0.1cm
%---------------------------------------------------------------------------
In this section we study the lowest lying bulk excited states  corresponding to different quantum number sectors $(S_z,N)$ with $S_z$ the total z-component spin of the state and $N$ the particle number.   Our main interest is in the energy gaps, defined as the difference from the ground state energy of the lowest-lying energies for given quantum numbers)  and their variation with system size and proximity to the transition.  For ease of computation we consider only YC3. 

The triplet spin gap is defined  as 
\begin{equation}
\Delta_{T} = E(1,N)-E(0,N),
\end{equation}
where both states have $N_{\rm el}=N$, the number of sites.
The single particle gap is defined as
\begin{equation}
\Delta_{1p} = E(1/2,N+1)+E(1/2,N-1)-2*E(0,N),
\end{equation}
and similarly, the singlet two particle gap is
\begin{equation}
\Delta_{2p} = E(0,N+2)+E(0,N-2)-2*E(0,N).
\end{equation}
These gaps all involve the lowest energy state in a given quantum number sector; in addition, we define the singlet gap $\Delta_S$ as the gap between the lowest two states with $N$ electrons and total spin $S=0$.  To exclude triplets with $S_z=0$, we add in $a S^2$, the total spin operator squared, into the Hamiltonian, with $a=5$,
raising all nonsinglets to high energy. Then successive singlets are obtained by adding to the Hamiltonian a projector based on the previous states found $\psi_i$: $H = H + \sum_i b |\psi_i\rangle\langle\psi_i|$ (with $b=1$ in most cases). 

%---------------------------------------------------------------------------
\begin{figure}[t]
\includegraphics[width=\linewidth]{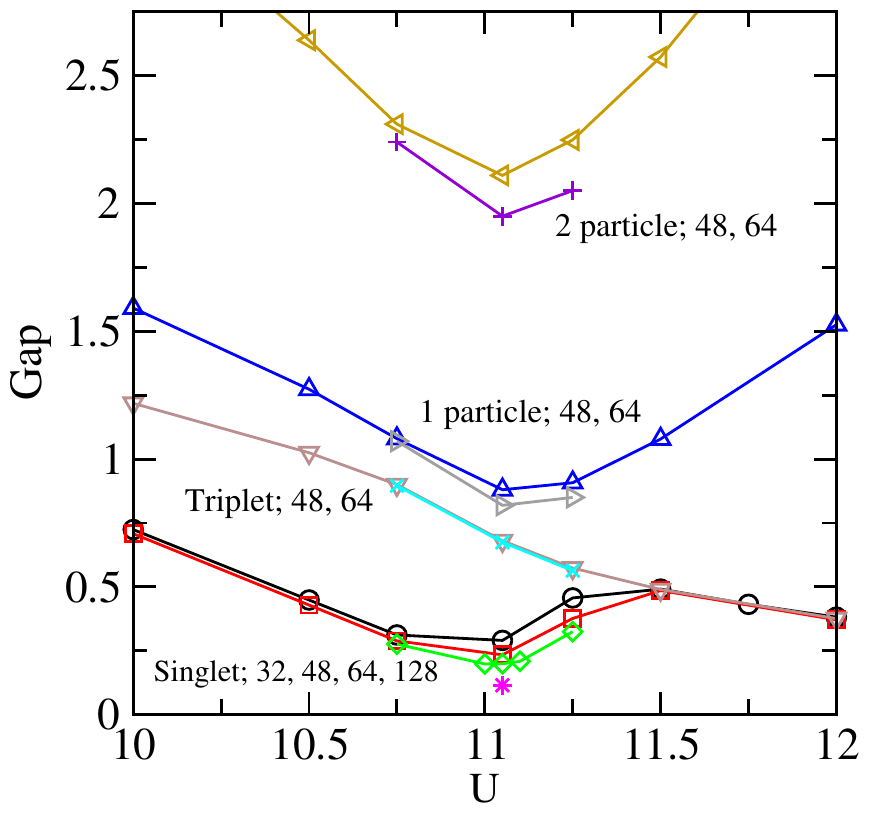}
\vskip -0.3cm
\caption{Various energy gaps for YC3 near the critical point.  The integers indicate the central sweep region $w$ for each curve,
where in all cases the curves for smaller $w$ are above those for larger $w$.
The total length in all cases is 128, except the $w=128$ point (pinkish star), where the length was 256.    DMRG convergence errors are smaller than the symbol size. The symbols at $U$ just greater than $11$ are at $11.05$, the estimated transition point. }
\vskip -0.4cm
\label{fig:Gaps}
\end{figure}
%---------------------------------------------------------------------------

It is important to exclude edge states in making these excited states.  We found that there are a number of edge excitations which are often lower in energy than bulk states.  Therefore, we used restricted-sweeping DMRG~\cite{Stoudenmire2012DMRG2D}, where the outer edges of the system are fixed in the ground state (by freezing the MPS tensors) and optimization occurs over a window of size $w$.

\begin{figure}[h]
\includegraphics[width=\linewidth]{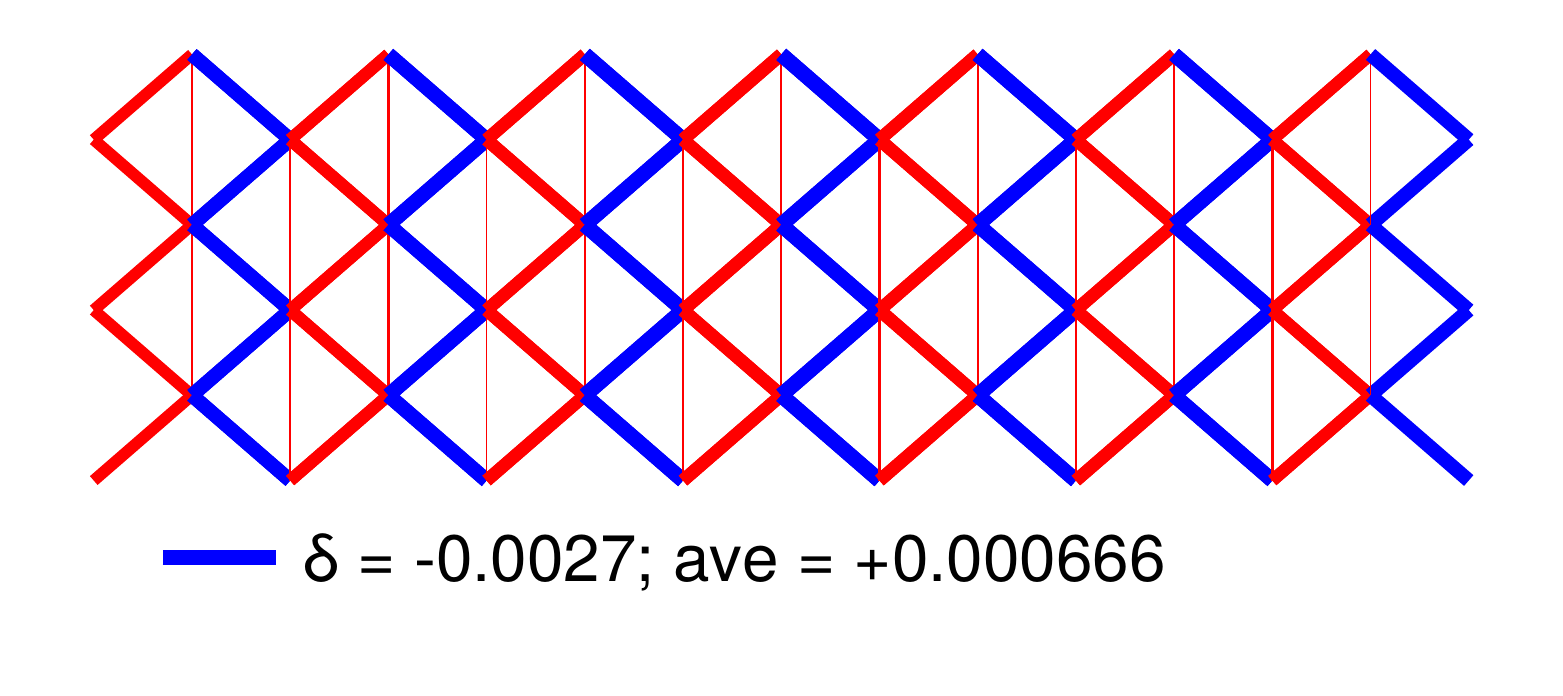}
\vskip -0.6cm
\caption{Off-diagonal measurement of $\langle \psi_1 | S^z_i S^z_j | \psi_0 \rangle$ for near neighbor bonds $i,j$ between the lowest energy and first excited singlet state for $U=11.05$, $w=64$, in the central part of the system. The thickness of the lines indicates the value, and the color of the bonds indicate the sign.  The pattern is consistent with a momentum $k_y=0$, $k_x=\pi$ excitation.}
\vskip -0.2cm
\label{fig:psi1SzSzpsi0}
\end{figure}

We find that all the gaps except the singlet remain open and substantial.  In contrast, the singlet gap is not only the smallest, it decreases significantly with $w$ near the transition. The results are consistent with the gap closing at the critical point (near $U=11.05$) in the $L_x\rightarrow \infty$ limit.     To extract the momentum and understand the nature of the singlet excitation, it is convenient to measure off-diagonal operators between the ground and excited states, as shown in Fig.~\ref{fig:psi1SzSzpsi0}.  The sign pattern in $\langle \psi_1 | S^z_i S^z_j | \psi_0 \rangle$ indicates a momentum $(\pi,0)$, with the nature of the excitation resembling the order parameter neutral order parameter $\mathcal{O}$. Note that this state is far in energy from any two particle (charge $2e$) states, as evidenced by the large two particle gap, and is non-degenerate. 

%---------------------------------------------------------------------------
%---------------------------------------------------------------------------

\begin{figure}[h]
\includegraphics[width=0.8\linewidth]{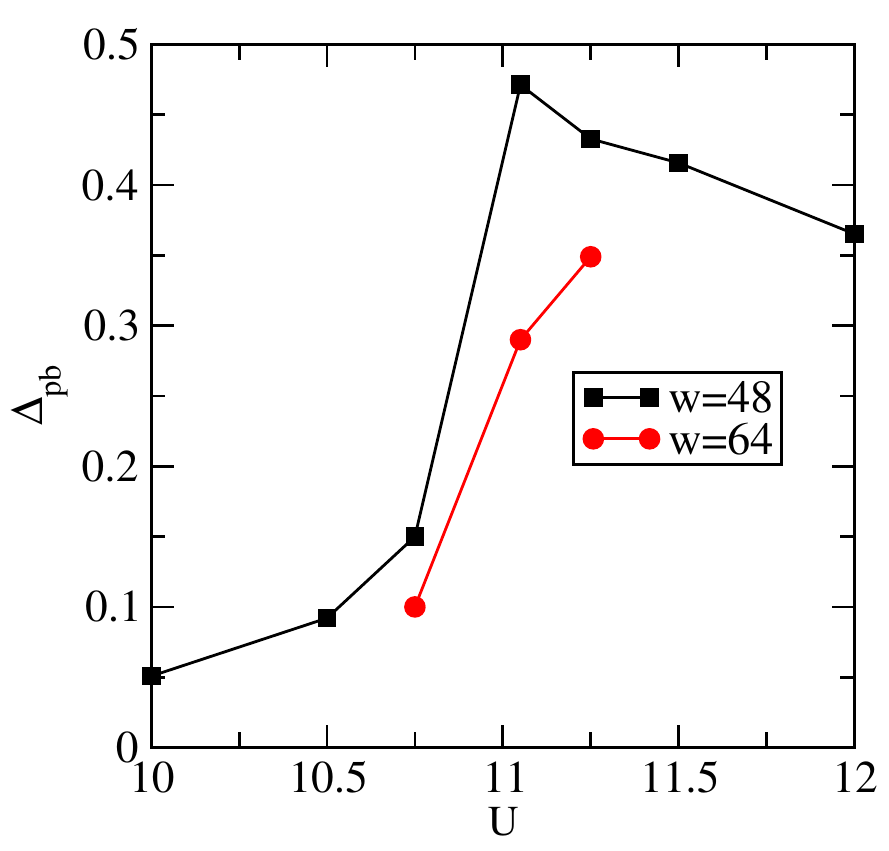}
\vskip -0.2cm
\caption{Pair binding $\Delta_{\rm pb}$ derived from the data shown in Fig.~\ref{fig:Gaps}.  }
\label{fig:PB}
\vskip -0.4cm
\end{figure}

It is interesting to look at the pair binding energy $\Delta_{\rm pb} \equiv \Delta_{\rm 2p} - 2 * \Delta_{\rm 1p}$, which is shown in Fig.~\ref{fig:PB}.  The positive values indicate a repulsion between the two particles, with the repulsion reduced on the larger system, and increasing with $U$.

%---------------------------------------------------------------------------
%---------------------------------------------------------------------------

%---------------------------------------------------------------------------
\section{Discussion and Conclusion} \label{sec:Section6}
%---------------------------------------------------------------------------
A long-standing goal of theoretical physics has been to identify and understand the properties of Hamiltonians giving rise to ``spin liquids'': insulating states of matter characterized by an absence of magnetic or other translation symmetry breaking long ranged order and the presence of entanglement and topological properties. One attractive candidate for spin liquid physics is the ``fluxed'' Hubbard model of electrons moving on a two dimensional lattice subject to an orbital (but not a Zeeman) magnetic field and to an on-site interaction. In different variants of this model spin liquid states have been proposed as intermediate coupling ground  states separating weakly correlated states with properties adiabatically connected to non-interacting particle ground states, and topologically trivial antiferromagnets occurring when the correlation strength becomes very large.

In this paper we have studied the Hubbard model on a triangular lattice with flux $\pi/2$ per plaquette. The non-bipartite nature of the triangular lattice Hubbard model acts to frustrate conventional magnetic order, widening the window where spin liquid behavior may potentially occur. The flux further introduces topology into the system. The small $U$ state is a $\nu=2$ (one for each spin) integer quantum Hall state, the large $U$ state is a $120^\circ$ antiferromagnet, and the two states are believed to be separated by an intermediate $U$ Kalmeyer-Laughlin chiral spin liquid state~\cite{Kalmeyer87}. Direct simulation of the two dimensional model is impractical. Following previous work~\cite{Knap2024,divic2024csl} we have used density matrix renormalization group methods to determine properties of small radius cylinders. Previous work  however used iDMRG methods; here we use finite system DMRG in which fixed length cylinders with open boundary conditions are studied, either with fixed values of the Hamiltonian parameters or under scan conditions~\cite{chepiga2021scan} in which the Hamiltonian parameters are varied smoothly over the length of the system. The finite system DMRG methods provide complementary information to the iDMRG studies. 

Our work confirms previous suggestions~\cite{Knap2024,divic2024csl} of a spin-gapped intermediate $U$ phase separated from the small $U$ integer quantum Hall phase by a $T=0$ phase transition and characterized (for the odd-radius cylinders we study) by a staggered order parameter noted by Divic et al. and corresponding to a bond-dimerized state. Use of the scan techniques confirms a continuous transition at the phase boundary, with exponents $0.06 < \beta < 0.35$ (best fit $0.14$) and $0.4 < \nu < 0.93$ (best fit $0.80$), extracted from the YC5 cylinders. However, as seen from Fig.~\ref{fig:ScanCollapse}(e,f) the scan data are consistent with Ising critical exponents  $\beta=1/8$ and $\nu=1$. The critical behavior from the divergence of correlation lengths on the IQH side, measuring decay away from edges of both the dimerization order parameter and transverse edge currents in the YC5 cylinders non-scans suggest $\nu=0.99$. This determination of $\nu$ seems more accurate than the scaling collapse. An interesting question is whether the dimerized phase may be viewed as a finite-radius instability of the CSL; one indication that this may be the case is the rapid decrease of the dimerization order parameter amplitude with increasing radius.

Changes across the transition in our computed entanglement spectra reveal (as does previous work~\cite{Knap2024,divic2024csl}) a change in degeneracy structure consistent with the expected  topological transition; the identification of the intermediate $U$ phase as a spin liquid is confirmed by the presence of spin but not charge pumping and by the magnitude of the topological component of the entanglement entropy. The scan-DMRG also revealed an interesting phase slip/domain wall associated with the spatial location where interactions cross the critical value. 

These results provide strong additional evidence for a 1+1D second-order transition in the odd circumference  DMRG cylinders and support the broader picture of an intermediate-$U$ CSL phase arising from magnetic frustration and electron topology (encoded here by orbital flux). The finding (consistent with previous work) of spin but not charge pumping in the intermediate-$U$ phase  and the change in entanglement spectra across the transition are strong evidence in favor of this picture, implying that the transition at $U_c$ is a topological Mott transition.

However our calculation of excitation spectra on a small radius cylinder is in one important respect inconsistent with a previous picture~\cite{divic2025}  of this topological Mott transition, which argues that at  the transition separating IQH and CSL phases on a flux $\pi/2$ triangular lattice there should be a gapless triplet mode consisting of a $q=0$ charge neutral excitation  and two states derived from it by action of the generators of the SU(2) symmetry first identified by Yang~\cite{Yang89}. The existence of a gapless or at least relatively soft charge $2e$ mode is also suggested by very recent reports of topological chiral superconductivity in the lightly doped version of the model we study~\cite{Pichler25,Chen25,Kuhlenkamp25}. For YC3, we find strong evidence that just one mode becomes gapless at the transition,  that this mode is the momentum $-\pi$ dimerization fluctuation associated with the (classically definable) order parameter introduced by Divic et al~\cite{divic2024csl}, and that all charge $2e$ excitations remain very high in energy.

The discrepancy may arise from finite-size effects. Our excitation results are on YC3. Even in this narrow width one sees a depression of the (large) single and two particle gaps near the transition, although it is important to note that the pair interaction energy (Fig.~\ref{fig:PB}) is repulsive.  It may be that one or both of these gaps close as $L_y \to \infty$. We attempted calculations on YC$5$ but could not obtain sufficiently converged data near the critical point to perform a reliable $1/L_y$ extrapolation. Alternatively the topological Mott transition may not require   gapless long-wavelength charge neutral and $2e$ modes.   As discussed in section~\ref{sec:Section3B}, the entanglement spectrum and spin pumping at $U>U_c$ may be understood in terms of the dimerized spin model expected in a Mott-localized phase; these phenomena are also observed in even-radius cylinders where $U_c$ seems to mark a crossover, not a transition~\cite{Knap2024}. Possibly Mott localization may change the structure of the boundary models without explicitly closing a bulk gap.  In approximate analyses of related models~\cite{Wagner24} it was argued that the change in topology could arise from a change in bulk properties of Green function zeros rather than from opening and closing of physical gaps. Our finding may also be related to  theoretical studies of ``symmetric mass generation'' where changing an interaction may gap a boundary fermion without closing and reopening a bulk gap~\cite{Wang22}.  

Investigation of the structure of the phase transition  and order parameter magnitude, and of the excitation spectrum for wider radius cylinders is of great interest and importance, as is deeper understanding of the superconductivity and other properties of  doped systems. 

% ============================================================================================
\section*{Acknowledgments}
\vskip -0.5cm
CAG was supported by the Graduate Fellowship from Eddleman Quantum Institute at UC Irvine. RMM and SRW were supported by the U.S. NSF under Grant
DMR‑2412638. AJM was supported  by Programmable Quantum Materials, an Energy Frontier Research Center funded by the U.S. Department of Energy (DOE), Office of Science,
Basic Energy Sciences (BES), under award DESC0019443. We thank V. Crepel, S. Divic, M. Stoudenmire, and T. Scaffidi for helpful discussions.  The calculations were performed using the ITensor library~\cite{itensor}. The Flatiron Institute is a division of the Simons Foundation.
% ============================================================================================

% ============================================================================================
\vskip -0.5cm
\section*{Data availability}
\vskip -0.5cm
The data that support the findings of this article are openly
available~\cite{data-availability}.
% ============================================================================================

\bibliography{refs}
\newpage 
\appendix

% ==============================================================================
%---------------------------------------------------------------------------
\onecolumngrid
%---------------------------------------------------------------------------
\begin{center}
\ \vskip -0.1cm
{\large\bf Appendix}
\end{center}
%---------------------------------------------------------------------------
\vskip 0.2cm \
%---------------------------------------------------------------------------
\twocolumngrid
%---------------------------------------------------------------------------

%%%%%%%%%%%%%%%%%%%%%%%%%%%%%%%%%%%%%%%%%%%%%%%%%%
\renewcommand{\theequation}{A\arabic{equation}}
%%%%%%%%%%%%%%%%%%%%%%%%%%%%%%%%%%%%%%%%%%%%%%%%%%
\setcounter{equation}{0}
% ==============================================================================

%---------------------------------------------------------------------------
\section{Scans extra data}
%---------------------------------------------------------------------------
\vskip -0.2cm

%---------------------------------------------------------------------------
\begin{figure}[b]
\includegraphics[width=\linewidth]{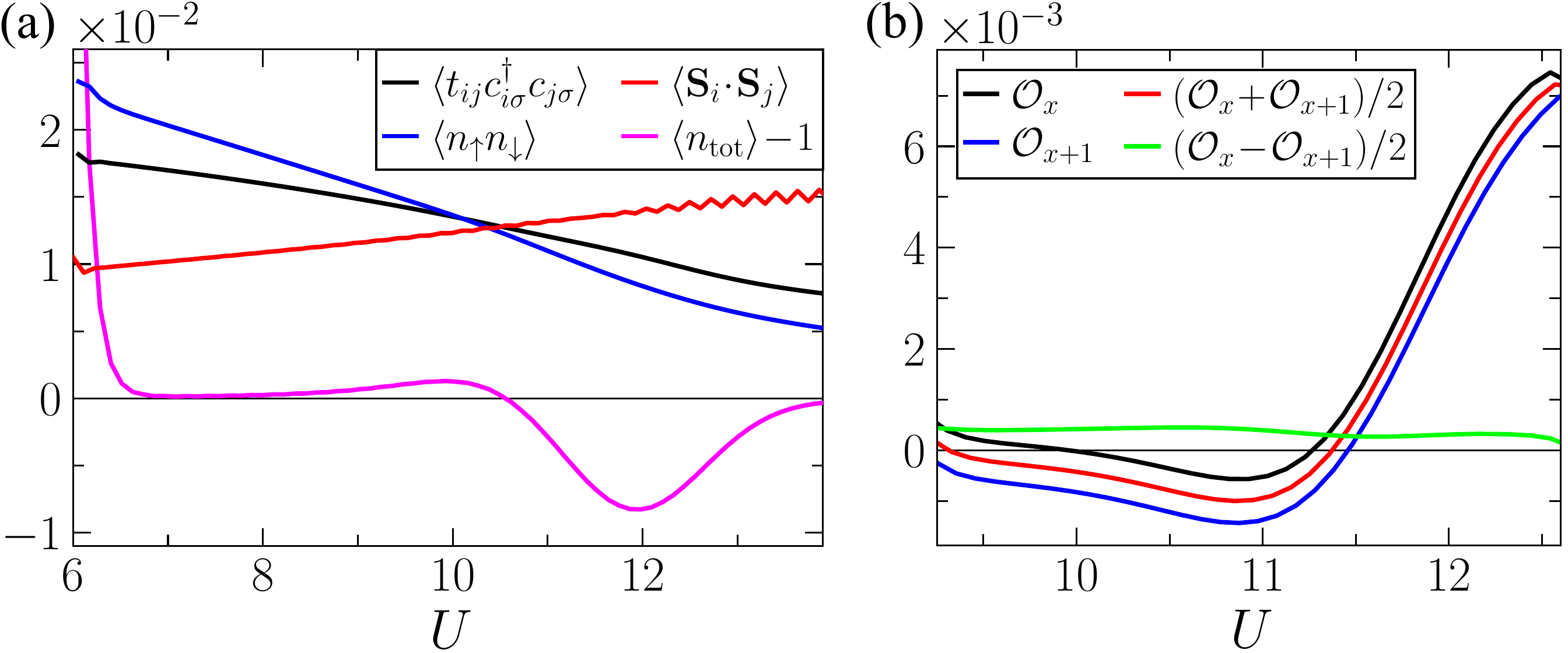}
\vskip -0.2cm
\caption{(a) Additional results for DMRG scans on a YC$5$ cylinder. The black, blue, red, and magenta lines are proportional to the local kinetic energy, on-site interaction energy, spin-spin correlation on the diagonal bonds, and the deviation of the charge density from half-filing, respectively. (b) The order parameter $\mathcal{O}_x$ from a DMRG scan. The black and blue solid lines are the order parameter measured on the even and odd columns, respectively, with the red and green line their difference; see the text.}
\label{fig:ScansExtraData}
\end{figure}
%---------------------------------------------------------------------------

In the scans, several measurements can be performed to provide additional insights. For instance, one could ask how the local kinetic and interaction energies, proportional to the hopping amplitudes $t_{ij}c_{i\sigma }^\dagger c_{j\sigma }$ and the on-site repulsion $U n_{\uparrow}n_{\downarrow}$, respectively, change as a function of $U$. Our Fig.~\ref{fig:ScansExtraData}(a) shows both the local kinetic energy, shifted by $10$, and the Hubbard interaction energy as black and blue solid lines, respectively. Note that if one adds the two curves and subtracts the shift of $10$, one obtains the total local energy. The energies vary  smoothly with $U$, which they would not if the transition were first order. Note that the on-site interaction energy has a decrease in slope near the transition, indicating that the CSL, which is a Mott phase, is more effective in reducing the on-site Coulomb repulsion energy. 

In addition, Fig.~\ref{fig:ScansExtraData}(a) shows the diagonal spin–spin correlations (red line) measured on every column and multiplied by a constant to make all the scales comparable. One observes that the spin correlations increase almost linearly with $U$, but after the transition there is a clear alternation, which is what the dimerization order parameter measures. The magenta line shows the deviation of the local charge density from half-filling (also multiplied by a constant for visibility).  The IQH phase has localized excess charge on its ends. The excess charge on the left is highly localized at small $U$, whereas the excess on the right near $U_c$ is broadened by the continuous transition. 

Fig.~\ref{fig:ScansExtraData}(b) shows in more detail how we calculate the order parameter $\mathcal{O}_x$ in the scans. The black and blue lines are the order parameter measured on the even and odd columns, respectively. Their average (red line) is the value reported in our results in the main text, whereas their difference (green line) corresponds to the oscillation due to the finite gradient $\delta U$ used in the scan.

%---------------------------------------------------------------------------
\begin{figure}[t]
\includegraphics[width=\linewidth]{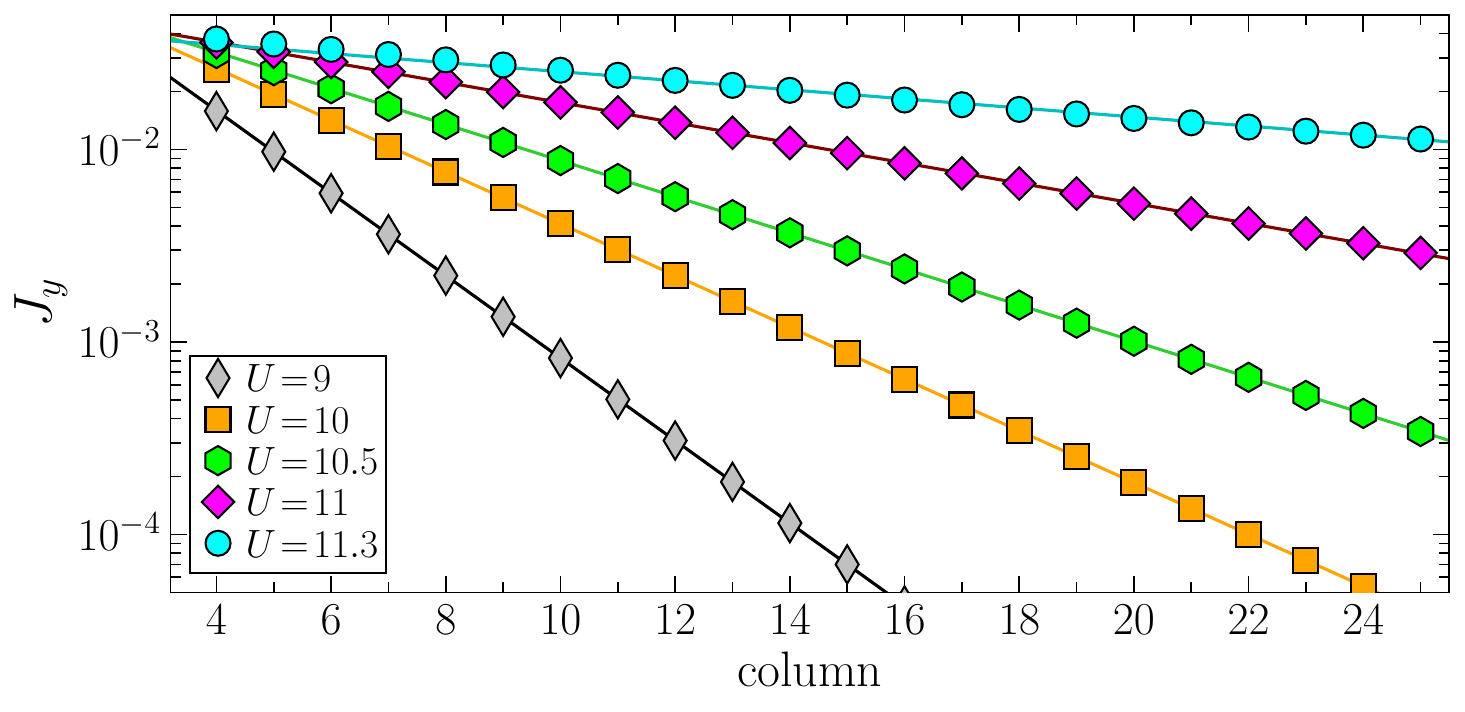}
\vskip -0.4cm
\caption{(a) Decay of the vertical currents $J_y$ as a function of distance from the edge in $80\!\times\!5$ YC cylinders, for different values of $U$. Solid straight lines in the semi‑log plot indicate the exponential fits used to extract the correlation length for each $U$.}
\label{fig:Currents}
\vskip -0.4cm
\end{figure}
%---------------------------------------------------------------------------

\vskip -0.3cm
%---------------------------------------------------------------------------
\section{Electron currents}
%---------------------------------------------------------------------------
\vskip -0.2cm

Our Fig.~\ref{fig:Currents} shows spatial dependence of the transverse currents $J_y$ (along the $y$-direction) as we move away from the edge. A fit of the exponential decay into the bulk (straight line in the semi-log plot) yield the same correlation length $\xi(U)$ as obtained from the order‑parameter profiles in the main text.

%---------------------------------------------------------------------------
\section{Entanglement spectrum procedure}
%---------------------------------------------------------------------------

Here we provide the technical details of the entanglement spectrum construction, referred to in Sec. \ref{sec:Section3B} of the main text.

The ground state on a YC–5 cylinder of length $L_x=80$ is obtained with finite‑DMRG, keeping a bond dimension $\chi=2000$.  The wavefunction is written as an MPS in snake ordering, where one ring comprises the five physical sites at fixed $x$.  Placing a bipartition between rings $x=\ell$ and $x=\ell+1$, we move the orthogonality center immediately to the left of the cut and perform a singular‑value decomposition (SVD)
\begin{equation}
  \psi_{(\alpha\,\sigma_{\ell})\,(\beta\,\sigma_{\ell+1})}
      =\sum_{\gamma=1}^{\chi}
        U_{(\alpha\,\sigma_{\ell}),\gamma}\,
        \lambda_{\gamma}\,
        V^{\dagger}_{\gamma,(\beta\,\sigma_{\ell+1})},
  \label{eq:SVD_detailed}
\end{equation}
where $\alpha$ ($\beta$) is the index of the left (right) bond and $\sigma_{\ell}$, $\sigma_{\ell+1}$ are the physical indices of the two sites adjacent to the cut.  The fusion of $(\alpha,\sigma_{\ell})$ and $(\beta,\sigma_{\ell+1})$ produces composite indices of dimension $\chi d$ with $d=4$.  The resulting singular values $\lambda_{\gamma}$ define the Schmidt probabilities $\lambda_{\gamma}^{2}$ and the entanglement energies $\varepsilon_{\gamma}=-\ln\lambda_{\gamma}^{2}$.

%---------------------------------------------------------------------------
\begin{figure}[t]
\centering
\includegraphics[width=.95\linewidth]{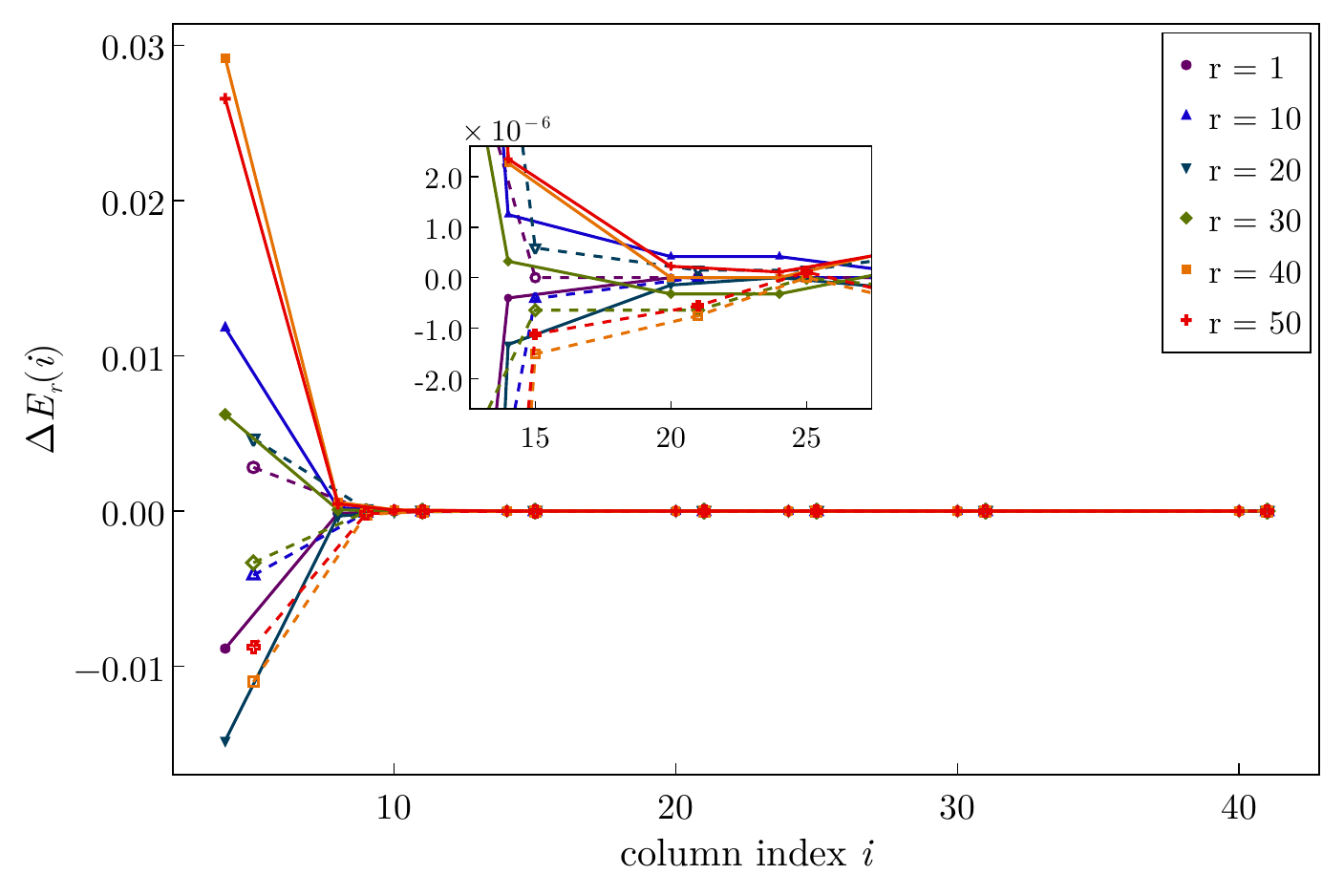}
\caption{Stability of the entanglement spectrum (ES) under off-center bipartitions. We show \(\Delta E_{r}(i)=E_{r}(i)-E_{r}(i_0)\) with \(E_{r}(i)=-\ln\lambda_{r}^{2}(i)\) and \(i_0\) set at column 40 for a representative interaction \(U=6\) on a cylinder of length \(L_x=80\) (400 sites). Curves correspond to several fixed ES ranks \(r\); for each rank the color/marker is fixed, with \emph{even} cuts drawn as solid lines with filled markers and \emph{odd} cuts as dashed lines with open markers. The inset zooms the vicinity of \(i\!\approx\!100\), \(\Delta E\!\approx\!0\). Deviations between \(i \simeq 20\) and the center remain at the order of $10^{-6}$, justifying the off-center choice used in our ES calculations to save compute time without measurable bias.}
\label{fig:es-cut-profile}
\end{figure}
%---------------------------------------------------------------------------

After the SVD we assign a transverse momentum $k_y$ to each Schmidt level $\gamma$ as follows.

\emph{(i) Isolate a single Schmidt vector on the left.}
Because the MPS is in mixed–canonical form with the orthogonality center at the cut, selecting one value of the bond index $\gamma$ on $U$ (one ``column" of $U$) defines a normalized left Schmidt state $|L_\gamma\rangle$ in the
form of a ``half-MPS"--just an MPS of length $\ell$.

\emph{(ii) Apply the translation $T_y$ to determine the momentum.}
Let $(x,y)$ label the $y=1,\dots,5$ physical sites of ring $x$ in snake ordering. Let $S_{i,j}$ be the fermionic SWAP that exchanges sites $i$ and $j$. On a single YC–5 ring (“column”) we implement one-site translation with four SWAPs,
\begin{equation}
T_y^{(x)} \;=\; S_{(x,5),(x,4)}\; S_{(x,4),(x,3)}\; S_{(x,4),(x,2)}\; S_{(x,2),(x,1)} ,
\label{eq:Ty_column}
\end{equation}
which realizes the cyclic permutation $1\!\to\!2\!\to\!3\!\to\!4\!\to\!5\!\to\!1$. Along the snake, each swap is near-neighbor, making this form particularly easy to apply. The translation on the left half-cylinder factorizes ring-by-ring,
\begin{equation}
T_y \;=\; \prod_{x\le \ell} T_y^{(x)} .
\label{eq:Ty_product}
\end{equation}
We apply $T_y$ \emph{column after column}, progressively updating the left-half MPS, forming the state $\langle \tilde L_\gamma\rangle$ and then measure the overlap
\begin{equation}
\langle L_\gamma|\tilde L_\gamma\rangle \;=\;
\langle L_\gamma|\,T_y\,|L_\gamma\rangle \;=\; e^{i\,k_{y,\gamma}}.
\label{eq:overlap_phase}
\end{equation}
 
Because of the five-fold rotational symmetry of YC–5, the allowed momenta are quantized as
\begin{equation}
k_{y,\gamma}\in\Big\{0,\ \pm\tfrac{2\pi}{5},\ \pm\tfrac{4\pi}{5}\Big\}.
\label{eq:ky_quantized}
\end{equation}

We benchmark the dependence of the entanglement spectrum (ES) on the longitudinal
position of the bipartition.  Let the ES levels be indexed by rank \(r\) and column
index \(i\), with
\begin{equation}
  E_{r}(i) \equiv -\ln \lambda_{r}^{2}(i), \qquad i_0 = L_x/2L_y .
\end{equation}
To reduce the cost of the momentum labeling (the translation operator acts only on the
shorter half-cylinder), we evaluate the ES at \(i \simeq L_x/4L_y\) instead of the center
\(i_0\).  Figure~\ref{fig:es-cut-profile} plots
\begin{equation}
  \Delta E_{r}(i) \equiv E_{r}(i) - E_{r}(i_0)
\end{equation}
versus the column index \(i\) for several fixed ranks \(r\).  For each rank we keep a single color/marker and separate even and odd cuts: even cuts are solid lines with filled markers, odd cuts are dashed lines with open markers.  Once the cut is sufficiently far from the left edge, the deviation between \(i \simeq L_x/4L_y\) and the center remains at the level of \(\mathcal{O}(10^{-6})\).  This establishes that taking ES at \(i \simeq L_x/4L_y\) is center-equivalent within our numerical precision while substantially reducing the computational cost of the momentum-labeling step.

We also assess convergence of the entanglement spectrum with bond dimension \(\chi\) at that cut. For fixed Schmidt ranks \(r\) we compare the entanglement energies \(E_r(\chi) \equiv -\ln \lambda_r^2(\chi)\) and define the deviation from the largest bond dimension \(\chi_{\mathrm{ref}}\) (here \(\chi_{\mathrm{ref}}=2000\)) as
\begin{equation}
  \Delta E_r(\chi) \equiv E_r(\chi) - E_r(\chi_{\mathrm{ref}}).
\end{equation}

\begin{figure}[t]
\centering
\includegraphics[width=.95\linewidth]{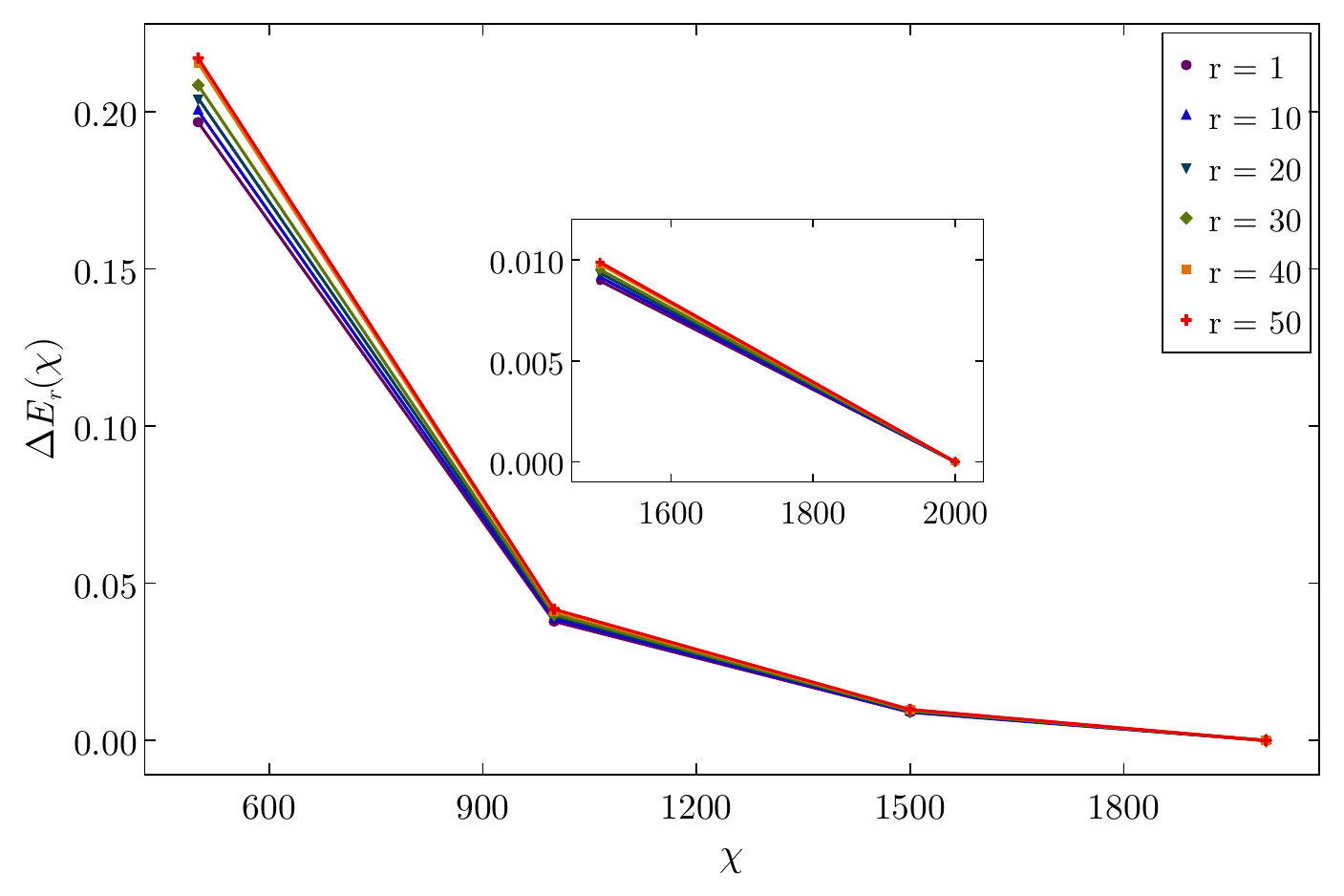}
\caption{Bond-dimension convergence of the entanglement spectrum at the center cut. The deviations \(\Delta E_r(\chi)=E_r(\chi)-E_r(\chi_{\mathrm{ref}})\) with \(E_r=-\ln\lambda_r^2\) and \(\chi_{\mathrm{ref}}=2000\) are shown for selected ranks \(r\). Data are for \(U/t=6\) and \(L_x=80\) with the cut at \(i_0=100\). \emph{Inset:} boxed zoom around \(\chi\in\{1500,2000\}\), highlighting the changes in the values from \(\chi=1500\) to \(\chi=2000\) are smaller than $10^{-2}$ for all ranks shown.}
\label{fig:bd_convergence}
\end{figure}

As Fig.~\ref{fig:bd_convergence} shows, \(\Delta E_r(\chi)\) decreases with \(\chi\) and, in particular, satisfies \(|\Delta E_r|<10^{-2}\) already at \(\chi=1500\) for all displayed ranks. Together with the off-center–cut test, this demonstrates that our ES calculations are well converged at moderate \(\chi\) and that measuring away from the center can be used to reduce cost without introducing bias.

%---------------------------------------------------------------------------
%---------------------------------------------------------------------------

%---------------------------------------------------------------------------
\section{Entanglement entropy extrapolations}
%---------------------------------------------------------------------------

To obtain the extrapolated von Neumann entropy $S$ on the YC3 cylinders, because we were concerned with observing divergences as we went through the critical point with long cylinder lengths, we tried to carefully optimize the extrapolations. We analyzed the  dependence of $S$ on the bond dimension $\chi$. We fit the data to a linear form in $1/\chi^p$, with the exponent $p$ optimized separately for each dataset. This procedure allows for a more flexible description of the $\chi$-dependence and reduces systematic bias.

\begin{figure}[t]
\includegraphics[width=.95\linewidth]{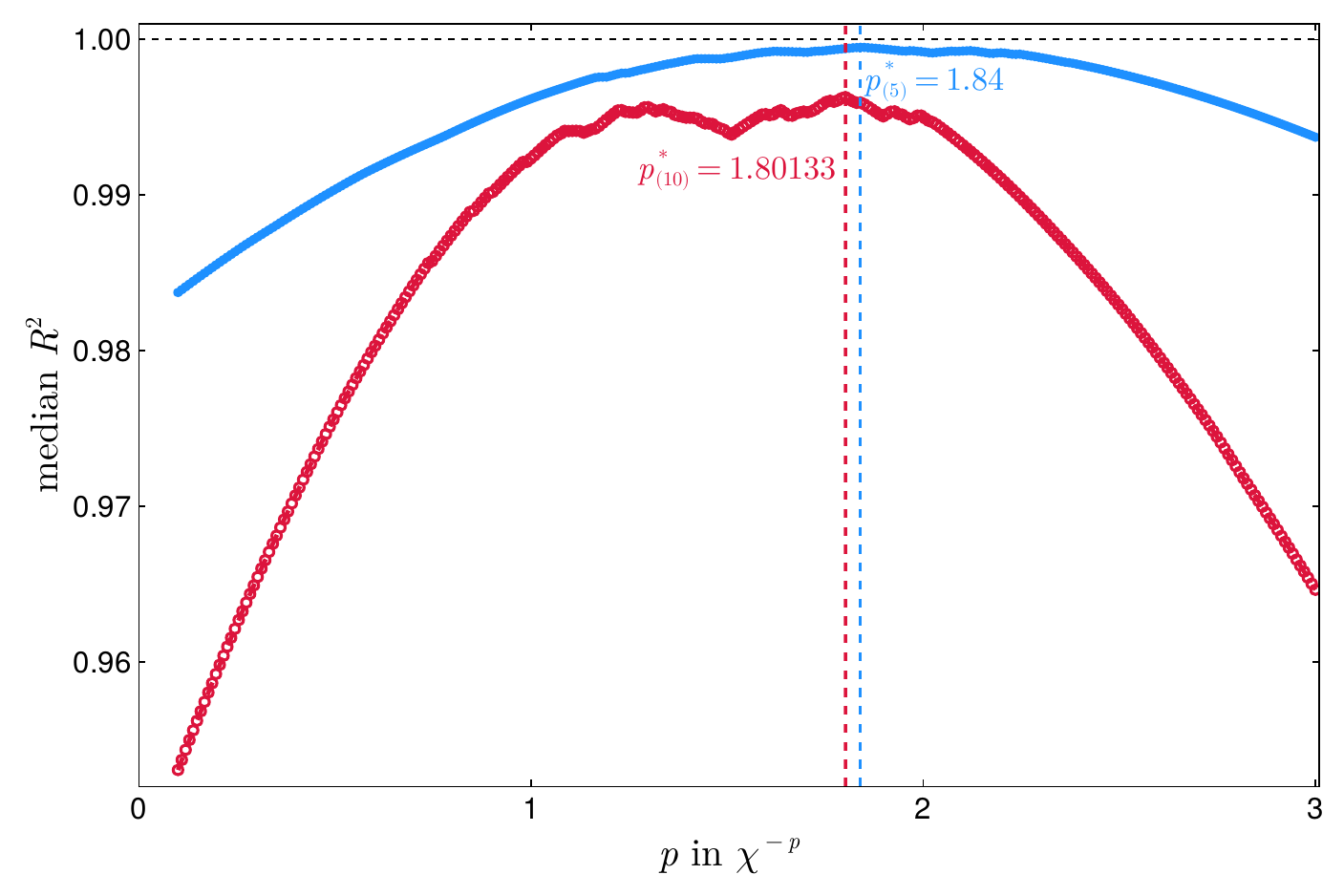}
\caption{Median coefficient of determination 
$R^2$ for the linear fits used to extrapolate the von Neumann entropy 
$S$ as a function of bond dimension $\chi$. Fits are performed in 
$1/\chi^p$, with the exponent 
$p$ optimized independently for each dataset. We compare extrapolations including the last five (blue) and the last ten (red) $\chi$ values. The consistently high $R^2$ values demonstrate the robustness of the extrapolation procedure.}
\label{fig:Rsquare}
\end{figure}

The stability of the extrapolation can be assessed by performing fits over two different windows: using the last five values of $\chi$, and using the last ten. In Fig.~\ref{fig:Rsquare} we plot the median coefficient of determination $R^2$ (a standard statistical measure telling how well the extrapolation is working), across all system lengths and interaction strengths for each choice of fitting window. In both cases the median $R^2$ remains very close to unity, indicating that a linear form in 
$1/\chi^p$ is generally appropriate. Near the phase transition, however, the data often exhibit mild residual curvature when more points at smaller $\chi$ are included. In this regime the ten–point window, while still producing high $R^2$, can visibly bend away from strict linearity, leading to a slight bias in the extrapolated intercept. The five–point window, restricted to the largest $\chi$ values, typically remains more linear and provides a more conservative estimate.

%---------------------------------------------------------------------------
%---------------------------------------------------------------------------

\begin{figure}[h!]
\includegraphics[width=.95\linewidth]{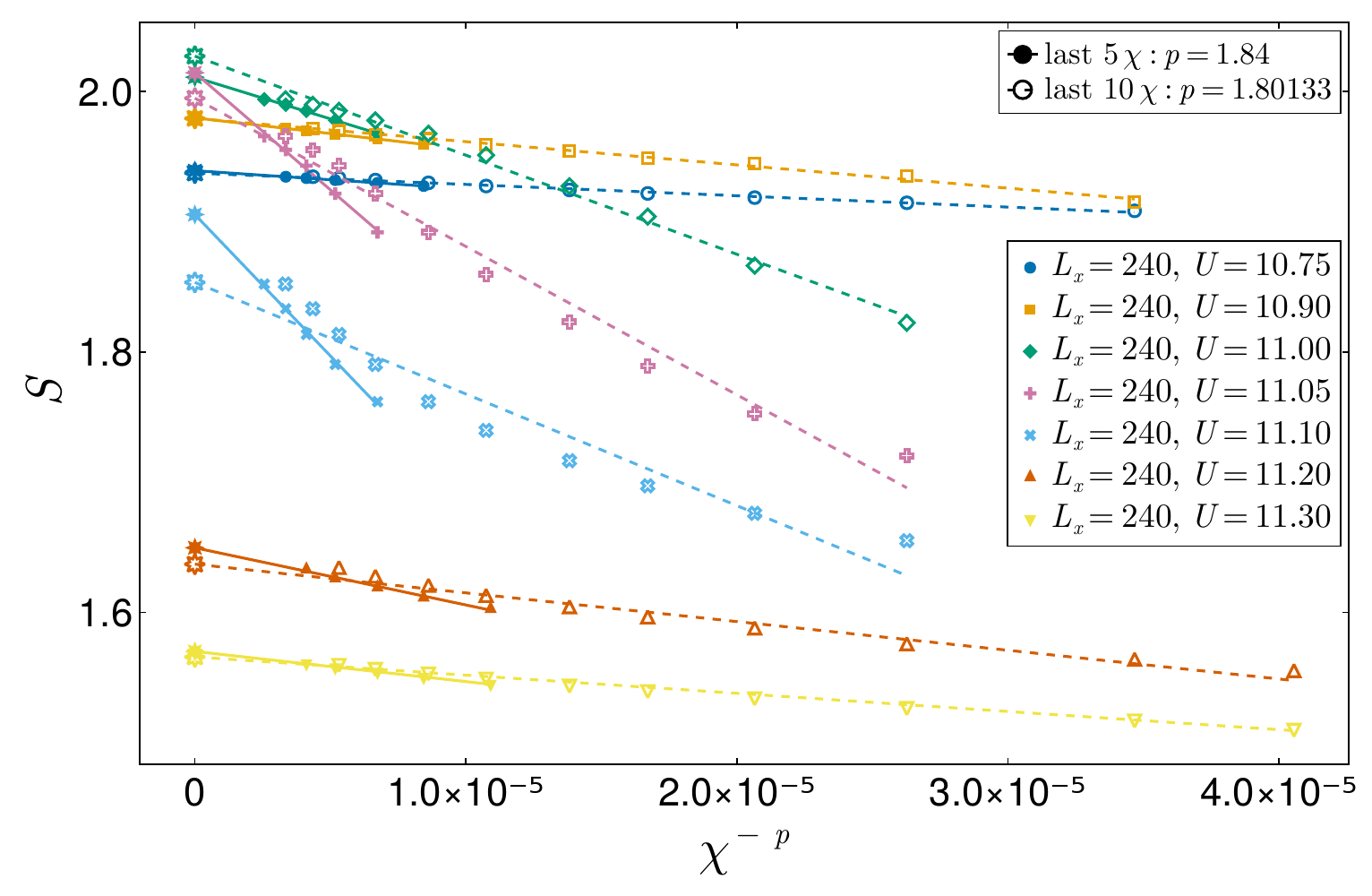}
\caption{Representative fits of $S$ versus $1/\chi^{p_{\mathrm{opt}}}$ at selected interaction strengths $U$ for fixed cylinder length $L_x$. Solid lines with filled markers use the last five $\chi$ values; dashed lines with open markers use the last ten. Lines are best linear fits in $1/\chi^p$ with $p=p_{\mathrm{opt}}$ optimized per dataset; the symbol at $1/\chi^{p_{\mathrm{opt}}}=0$ marks the extrapolated $S(\chi\to\infty)$. Near the transition the ten–point window can display mild curvature (despite high $R^2$), whereas the five–point window remains closer to linear, yielding a more conservative extrapolation.}
\label{fig:SvN_extrapolations}
\end{figure}

Figure~\ref{fig:SvN_extrapolations} illustrates these trends by overlaying the two windows at selected $U$ for fixed $L_x$: solid lines with filled markers (last five) track the linear behavior closely, while dashed lines with open markers (last ten) sometimes show curvature near the transition. We therefore report both windows and use their spread as an internal consistency check for the extrapolated values.

%---------------------------------------------------------------------------
%---------------------------------------------------------------------------

\begin{figure}[h!]
\includegraphics[width=\linewidth]{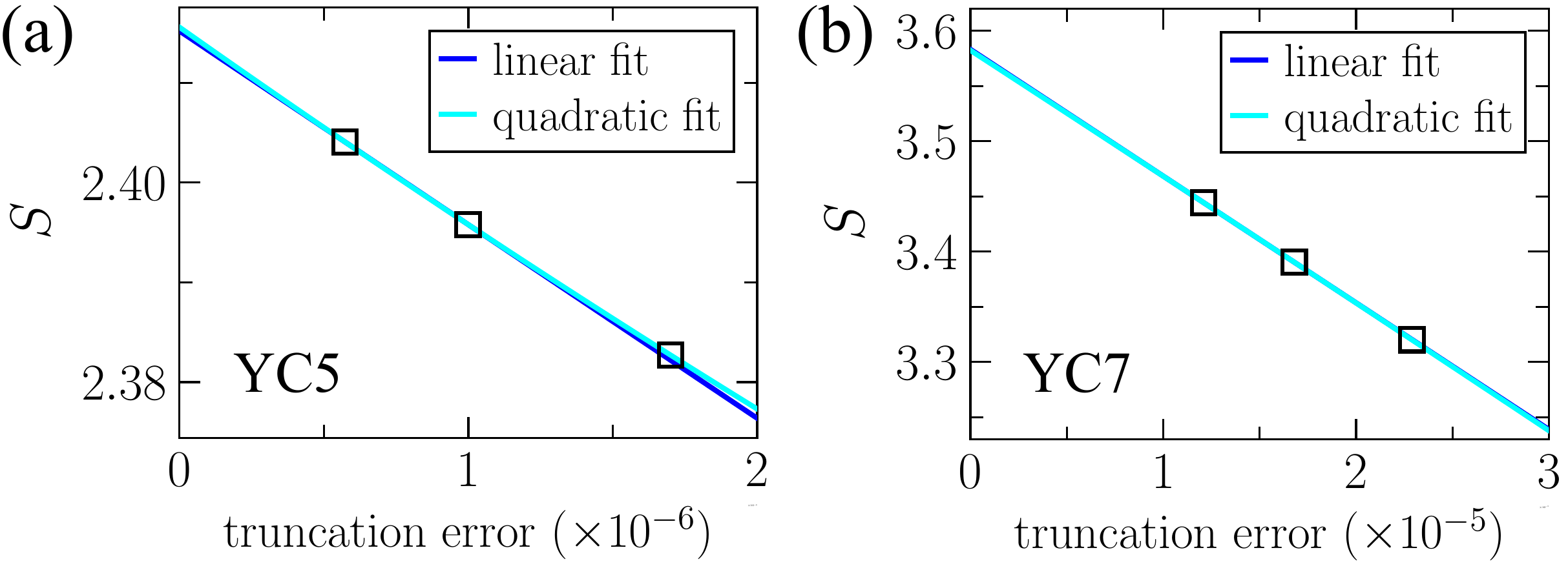}
\caption{Linear and quadratic extrapolations of the entanglement entropy in the YC$5$ and YC$7$ cylinders.}
\label{fig:S_errors}
\end{figure}

We now discuss additional details of the entanglement entropy data used in Fig.~\ref{fig:SvsL} to extrapolate the topological entanglement entropy. In this case the calculations were away from the critical point with relatively short cylinders, and linear extrapolations of $S$ with the truncation error seemed appropriate. Our Figure~\ref{fig:S_errors} shows the somewhat unexpected linear dependence of $S$ in the YC5 and YC7 cylinders as a function of the truncation error. We tested different power-law extrapolations with the truncation error, such as a linear fit of the data from the last two sweeps (blue lines) and a quadratic fit of the last three (cyan lines). Both approaches yield almost the same extrapolated $S$ within differences $\sim 10^{-4}$, so we do not provide error estimates for the results. It is worth noting, however, that the extrapolated $S$ in the YC7 cylinder differs by $10\%$ from the value in the last sweep at $\chi=16000$ despite the seemingly small truncation error $\sim 10^{-5}$. This should be contrasted with the YC3 and YC5 cylinders, where the corresponding differences are within $1\%$ and $0.1\%$, respectively.

%---------------------------------------------------------------------------
%---------------------------------------------------------------------------

\end{document}